\documentclass[a4paper,11pt]{article}
\pdfoutput=1 

\usepackage{jheppub} 

\usepackage[T1]{fontenc} 
\usepackage{subfig}
\usepackage{float}
\usepackage{bbold}
\usepackage[dvipsnames]{xcolor}

\usepackage{graphicx}

\allowdisplaybreaks[1]

\title{Two interacting scalars system in curved spacetime -- vacuum stability from the curved spacetime
Effective Field Theory (cEFT) perspective}

\author{Zygmunt Lalak,}
\author{Anna Nakonieczna}
\author[1]{and {\L}ukasz Nakonieczny \note{Corresponding author.}}

\affiliation{Institute of Theoretical Physics, Faculty of Physics, University of Warsaw \protect \\
ul.~Pasteura 5,~02-093 Warszawa, Poland }

\emailAdd{Zygmunt.Lalak@fuw.edu.pl}
\emailAdd{Anna.Nakonieczna@fuw.edu.pl}
\emailAdd{Lukasz.Nakonieczny@fuw.edu.pl}

\abstract{ 
In this article we investigated the influence of the gravity mediated higher dimensional operators on the issue of vacuum stability
in a model containing two interacting scalar fields. As a framework we used the curved spacetime Effective Field Theory (cEFT)
applied to the aforementioned system in which one of the scalars is heavy. After integrating out the heavy scalar we used the standard 
Euclidean approach to the obtained cEFT.  Apart from analyzing the influence of standard operators like the non-minimal coupling to gravity
and the dimension six contribution to the scalar field potential, we also investigated the rarely discussed dimension six contribution 
to the kinetic term  and the new gravity mediated contribution to the scalar quartic self-interaction.  
}

\begin{document} 
\maketitle
\flushbottom

\section{Introduction}
\label{sec:intro}

The Effective Field Theory (EFT) approach turns out to be an immensely useful framework to parametrize and analyze the effects of newly 
observed (or yet not observed, but expected) phenomena  in many branches of physics. In the case of particle physics it is most often 
invoked to describe possible influence of the beyond Standard Model (BSM) particles on observed quantities that can be measured 
in experiments (see, for example, \cite{BUCHMULLER1986621,Grzadkowski2010,Dedes2018} and references therein). 
Recently, there has been a renewed influx of efforts at writing down the systematic way of 
producing EFT from a set of minimal assumptions about new particles sector 
\cite{Henning_Lu_Murayama_2016,Henning_Lu_Murayama_2016_2,Drozd_Ellis_Quevillon_You_2016,Ellis2017}.  
In an effort to extend this approach to curved spacetime the formalism of the curved spacetime EFT has been recently developed in 
\cite{Nakonieczny_2019}.

In the current article we applied this formalism to investigate the problem of the vacuum stability in the presence of
non-trivial gravitational field. The problem of the SM vacuum stability is one of the open questions in modern theoretical cosmology.
For many years it has been investigated using both the flat spacetime approach 
\cite{SHER1989273, EliasMiro_Espinosa_Guidice_Isidori_Riotto_Strumia_2012,Degrassi_2012} 
(see also citations therein) and taking into account
the gravity effect in the case when Higgs was coupled to gravity 
\cite{Espinosa:2015qea,Czerwinska_Lalak_Nakonieczny_2015,
Burda:2015yfa,Burda:2015isa,Burda:2016mou,Bounakis:2017fkv,HAN2019314}.
More recently, there was a renewed surge of interest in 
better understanding the vacuum stability in curved spacetime in the case when scalar field is coupled non-minimally to gravity
\cite{Herranen_Markkanen_Nurmi_Rajantie_2015,Herranen_Markkanen_Nurmi_Rajantie_2014,Czerwinska:2016fky,
PhysRevD.95.025008,10.3389/fspas.2018.00040} (for some interesting discussion pertaining to the influence of the higher 
dimensional operators which couple kinetic terms of the Higgs and inflaton fields see \cite{Fumagalli:2019ohr}). 
The aforementioned papers follow mainly calculational route pioneered by classical papers
\cite{Coleman:1977py,PhysRevD.16.1248,PhysRevD.21.3305}, however there are also some alternative approaches 
like for example 
\cite{PhysRevLett.117.231601,PhysRevD.61.103508,PhysRevD.101.045021}.
At this point it is prudent to note that beside the Coleman-De Luccia bounces in the case of curved spacetime there are also other
forms of instantons, the most noticeable Hawking-Moss one \cite{HAWKING198235}. Some more discussion on their properties and
interpretation may be found, for example, in \cite{PhysRevD.76.064003,PhysRevD.94.024044}. 

Modern particle physics lacks full understanding of the non-gravitational interactions that may be relevant at the high energy scale 
related to the question of vacuum stability. Most importantly we do not know what is the precise description of 
dark matter or inflationary sector and how they interact with ordinary matter. Although there are many 
models for dark matter or inflation that are in agreement with observations we do not have sufficient data 
to prove which one is actually realized in Nature. In this context we may consider the requirement of vacuum stability, or at least metastability with the lifetime longer than the age of the Universe, as a non-trivial constraint on the beyond Standard Model physics.
Having this in mind it seems reasonable to use the framework that was developed 
precisely for this type of problems, namely the Effective Field Theory or more accurately its curved spacetime counterpart. 

As was mentioned above, the main purpose of this article is to show that the application of cEFT to the problem of vacuum stability 
is, first of all, possible and, secondly, may lead to a discovery of new interesting phenomena. To this end, we employed a 
model containing two scalar fields interacting through a quartic term (Higgs portal type of the interaction).
Assuming that one of the fields is heavy we integrated it out and obtained an effective theory for the light scalar. 
The novel feature of our approach manifests itself in the gravity mediated operators that do not have its counterparts 
in flat spacetime. From this we proceeded by analyzing contributions of these operators to the lifetime of 
false vacuum. To better understand the mechanism that leads to the obtained results we made use of two observables.
One of them is the Ricci scalar and is related to geometry, while the second one is the Euclidean counterpart of the 
energy density and is related to the matter content of the discussed model.  

The rest of the article is structured as follows. Section \ref{sec:outline} contains a description of the scalar sector and derivation
of cEFT. It also contains derivation of equations of motion and description of the numerical procedure used in solving them.
In Section \ref{sec:results} we described and discussed the obtained results. Section \ref{sec:summary} contains the summary 
of our findings.

\section{Action and equations of motion}
\label{sec:outline}

We begin our investigations by writing down the action for the gravity--matter system. For the description of the gravity sector 
we use the standard Einstein-Hilbert action
\begin{align}
S_g &= \int \sqrt{-g}\ d^4 x \left [  \frac{1}{16 \pi G} R \right ],
\end{align}
where $R$ is the Ricci scalar and $G$ is the Newton constant. In what follows, we will be using the $(+++)$ sign convention of \cite{MTW_1973}, 
which includes the mostly plus convention for the metric tensor $(-,+,+,+)$. Our matter sector will be composed of two scalar fields, namely
$H$ that may represent the Higgs doublet and $X$ that may represent the real scalar singlet of heavy dark matter.
In the case when dark matter mass is large enough it falls under the WIMPZilla category. For some up to date 
discussion of this type of dark matter we refer the reader to \cite{PhysRevD.96.103540,ARCADI20201}
and citation therein.
The $UV$ action for this sector is
\begin{align}
\label{actionUV_XH}
S^{UV}_{m} &= \int \sqrt{-g} d^4 x \bigg(
- \frac{1}{2} d_{\mu} H^{\dagger} d^{\mu} H - V\left(|H|^2\right) - \xi_{H} R |H|^2 
+ \nonumber \\
&- \frac{1}{2} d_{\mu} X d^{\mu} X - \frac{1}{2} m_{X}^2 X^2 - \xi_X R X^2 - \frac{1}{2} \lambda_{HX} X^2 |H|^2
\bigg),
\end{align}
where the potential for the complex $H$ field may be given by $V\left(|H|^2\right) = \frac{1}{2} m_{H}^2 |H|^2 + \frac{\lambda_{H}}{4!} |H|^4$  and $d_{\mu}$ is a covariant derivative that may in general contain gauge fields parts.
For the case where $X$ represents the heavy scalar singlet with mass $m_{X}^2 > 0$ 
(we assume the following mass hierarchy: $m_{X}^2 \gg  |m_{H}^2|$), 
$d_{\mu}$ reduces to the standard covariant derivative in curved spacetime  $\nabla_{\mu}$.
For now we may disregard gauge fields of the Standard Model and therefore we may set $H$ to be a real scalar field, which also implies $d_{\mu} H = \nabla_{\mu} H$.
In (\ref{actionUV_XH}) we also included the non-minimal coupling of both scalars to gravity, the strength of these couplings is controlled 
by parameters $\xi_H$ and $\xi_X$.

The next step is to obtain cEFT for the scalar $H$ that will be valid at energy and curvature scales smaller than $m_{X}$. To this end,
we will follow \cite{Nakonieczny_2019} and integrate out the heavy $Z_{2}$-symmetric real scalar singlet $X$. Having done this we may write the action
functional for our cEFT as 
\begin{align}
S_{cEFT} = \int \sqrt{-g} d^4 x \bigg(
&- \frac{1}{2} d^{\mu} H^{\dagger} d^{\mu} H -  \frac{1}{2} \tilde{c}_{dHdH} d_{\mu}|H|^2 d^{\mu}|H|^2
- \tilde{c}_{GdHdH} G^{\mu \nu} d_{\mu} |H|^2 d_{\nu}|H|^2 + \nonumber \\ 
&- \xi_X R |H|^2 - \tilde{c}_{H} |H|^2 - \tilde{c}_{HH}|H|^4 - \tilde{c}_{6}|H|^6 - V\left(|H|^2\right)
\bigg),
\end{align}
where we defined the curvature dependent coefficients in the following manner:
\begin{align}
\label{cdHdH}
\tilde{c}_{dHdH} &=   \frac{\hbar}{(4 \pi)^2} \frac{\lambda_{HX}^2}{12 m_{X}^2} 
\left (1 + \frac{\left ( \xi_X - \frac{1}{10} \right )}{m_{X}^2} R  \right ), \\ 
\label{GdHdH}
\tilde{c}_{GdHdH} &=  - \frac{\hbar}{(4 \pi)^2} \frac{\lambda_{HX}^2}{360 m_{X}^4}, \\
\label{cH}
\tilde{c}_{H} &=  \frac{\hbar}{(4 \pi)^2} \bigg [ \frac{\lambda_{HX}}{12 m_{X}^2} \left ( 2 \xi_X - \frac{1}{6} \right)^2 R^2 + \nonumber \\
&+ \frac{\lambda_{HX}}{4 m_{X}^2}
\left ( 2\xi_X - \frac{1}{30} \right ) \square R 
- \frac{\lambda_{HX}}{270 m_{X}^2} \left ( \mathcal{K} - R_{\mu \nu} R^{\mu \nu} \right ) \bigg ], \\
\label{cHH}
\tilde{c}_{HH} &= \frac{\hbar}{(4 \pi)^2} \bigg [
\frac{\lambda_{HX}^2}{4 m_{X}^2} \left ( 2 \xi_X - \frac{1}{6} \right ) R 
- \frac{\lambda_{HX}^2}{8 m_X^4} \left ( 2 \xi_X - \frac{1}{6} \right )^2  R^2  + \nonumber \\ 
&- \frac{\lambda_{HX}^2}{720 m_{X}^4} \left ( \mathcal{K} - R_{\mu \nu} R^{\mu \nu} \right ) 
+ \frac{\lambda_{HX}^2}{m_{X}^4} \left ( - \frac{1}{4} \xi_X + \frac{1}{40} \right ) \square R 
- \frac{\lambda_{HX}^2}{90 m_{X}^4}  \nabla_{\mu} \nabla_{\nu} R^{\mu \nu} 
\bigg ], \\
\label{c6}
\tilde{c}_6 &=   \frac{\hbar}{(4 \pi)^2} \frac{\lambda_{HX}^3}{12 m_{X}^2}
\left (1 - \frac{\left ( 2 \xi_X - \frac{1}{6} \right )}{m_{X}^2} R \right ).
\end{align}
As we may see, the coefficients (\ref{cdHdH})--(\ref{c6}) were calculated up to terms of order $\mathcal{O}(\mathcal{R}^2)$, where 
$\mathcal{R}^2 = \left\{ R^2, R_{\mu \nu}R^{\mu \nu}, \mathcal{K} \equiv R_{\mu \nu \rho \sigma}R^{\mu \nu \rho \sigma} \right \}$
and $R_{\mu \nu}$ is the Ricci tensor and $R_{\mu \nu \rho \sigma}$ is the Riemann tensor. Yet, as was pointed out in \cite{Nakonieczny_2019}, in the case when the Ricci scalar is non-zero the dominant contributions will come from terms linear in $R$. Moreover, in what follows we will not discuss the curvature contribution to dimension six operators, that is we will set $c_{GdHdH} = 0$ and consider $c_{dHdH}$ and $c_6$ independent of the Ricci scalar. This is justified by the fact that curvature dependent parts of these coefficients are suppressed by additional powers of $m_{X}^2$ as compared to the terms proportional only to $\lambda_{HX}$. 
Let us also point out that some of the coefficients proportional to the Ricci scalar possess coefficients 
that vanish for $\xi_X = \frac{1}{6}$, yet since the massive scalar explicitly breaks the conformal symmetry in the heavy sector
this is not always the case. Moreover, we see that the coefficients of the terms proportional to $\mathcal{K}$ and 
$R_{\mu \nu} R^{\mu \nu}$ do not depend on $\xi_X$. At this point an interesting problem to consider would be to check whether,
assuming that the heavy sector is massless and non-minimally coupled to gravity, there is a trace of the conformal symmetry
of the heavy sector in higher dimensional operators coefficients.  Unfortunately, our method of obtaining cEFT depends on the
fact that the heavy sector fields are massive and therefore it cannot be applied to the aforementioned problem. 
It seems possible to surpass this obstacle by using the non-local representation of the heat kernel as discussed in \cite{Barvinsky_Vilkovisky_1990,barvinsky2009covariant,
Avramidi_1991, AVRAMIDI1998557, Barvinsky_Mukhanov_Nesterov_2003},
yet this is beyond the scope of the current article.     


To sum up, the action functional for both gravity and matter parts of the investigated system is
\begin{align}
\label{action_cEFT_h}
S_{cEFT} &= \int \sqrt{-g} d^4 x \bigg( \frac{1}{16\pi G } R - \frac{1}{2} \nabla_{\mu} h \nabla^{\mu} h - \frac{1}{2}m_{H}^2 h^2
 - \frac{\lambda_{H}}{4!} h^4 - a_3 h^3 + \nonumber \\
&- \frac{1}{2} c_{dHdH} \nabla_{\mu} h^2 \nabla^{\mu} h^2 - \xi_H R h^2 - c_{HH} R h^4 - c_{6} h^6 - c_0 \bigg).
\end{align}
In the above we specified the $H$ field to be a real scalar $h$ (this may represent the real component of the Higgs doublet)
and fixed its potential to be $V(|H|^2) \equiv V(h) = \frac{1}{2}m_{H}^2 h^2 + \frac{\lambda_{H}}{4!} h^4  + a_3 h^3 + c_0$.
Although this potential contains the term $h^3$ which is not present in the Higgs sector, we may think of it as a part of the $\lambda_{eff}(h)$
term that allows us to capture the essential (from the point of view of vacuum stability) feature of the Higgs effective potential, namely the presence 
of the second minimum at large energies. To be precise, we do not imply that this type of potential is accurate at capturing quantitative 
behavior of the Higgs vacuum, like for example the numeric value of lifetime, but it is good enough to capture at least some qualitative 
behaviors like an increase or decrease of lifetime due to higher order operators.
Moreover, in (\ref{action_cEFT_h}) we rewrite the action in such a way that various coefficients denoted $c_i$ are pure numbers 
which is signalled by the absence of tildes above them. 

To describe the gravitational field we need to specify an ansatz for the metric. 
Since the standard procedure in the case of investigations of vacuum stability is to compute a solution to the Euclidean equations of motion 
in the case when the metric possesses a $S^3$ topology we may write the line element as
\begin{align}
ds^2 = d \tau^2 + A^2(\tau) \Big [ d \psi^2 + \sin^2(\psi) \left( d \theta^2 + \sin^2(\theta) d \phi^2 \right) \Big ].
\end{align} 
For this parametrization of the metric the non-zero components of the Einstein tensor are given by
\begin{align}
G^{\tau}_{~ \tau} &= 3 \frac{ \dot{A}^2 - 1}{A^2}, \\
G^{\psi}_{~ \psi} &= \frac{ 2 A \ddot{A} + \dot{A}^2 - 1}{A^2},
\end{align}
where the overdot is defined as $\frac{d A(\tau)}{d \tau} \equiv \dot{A}$. Meanwhile, the Ricci scalar is 
\begin{align}
\label{Ricci_sc}
R = - 6 \frac{ A \ddot{A} + \dot{A}^2 - 1}{A^2}.
\end{align}
On the other hand, the Euclidean action for cEFT is given by (basically 
speaking the process of obtaining the Euclidean action in this case comes down to the fact that the exponential factor $e^{iS_{cEFT}}$ 
in the path integral goes to $e^{-S_{cEFT}^{Euc}}$)
\begin{align}
\label{cEFT_euc}
S^{Euc}_{cEFT} = \int \sqrt{g} d^4 x \bigg(
-\frac{1}{16 \pi G} R + f(h) R + \frac{1}{2} \nabla_{\mu} h \nabla^{\mu} h
+ \frac{1}{2} c_{dHdH} \nabla_{\mu} h^2 \nabla^{\mu} h^2 + V(h)
\bigg),
\end{align}
where $f(h) = \xi_H h^2 + c_{HH} h^4$ and $V(h) = \frac{1}{2} m_{H}^2 h^2 + a_3 h^3 + \frac{\lambda_{H}}{4!} h^4 + c_{6} h^6 + c_0$.
For the above action the relevant Einstein equations could be rewritten as 
\begin{align}
\label{eq_G00}
G^{\tau}_{~ \tau} = \frac{1}{\kappa(h)} \bar{T}^{\tau}_{~ \tau}, \\
\label{eq_G11}
G^{\psi}_{~ \psi} = \frac{1}{\kappa(h)} \bar{T}^{\psi}_{~ \psi}, \\
\end{align}
where 
\begin{align}
\label{kappah}
\kappa(h) &= \bar{M}_{Pl}^2 - 2 f(h), \\
\label{bTupup}
\bar{T}^{\mu \nu} &=   \nabla^{\mu} h \nabla^{\nu} h
- g^{\mu \nu} \left [ \frac{1}{2} g^{\rho \sigma} \nabla_{\rho} h \nabla_{\sigma} h  \right ] 
+ c_{dHdH} \nabla^{\mu} h^2 \nabla^{\nu} h^2
- g^{\mu \nu} \left [ \frac{1}{2} c_{dHdH} g^{\rho \sigma} \nabla_{\rho} h^2 \nabla_{\sigma} h^2  \right ]  + \nonumber \\
&- g^{\mu \nu} V(h) + 2 g^{\mu \nu} \square f(h) - 2\nabla^{\mu} \nabla^{\nu} f(h)
\end{align}
and $\overline{M}_{Pl}^2 \equiv \frac{1}{8 \pi G} $ is the reduced Planck mass. 
Meanwhile, the equation for the scale factor can be written as
\begin{align}
\label{eq_A}
\ddot{A} = \frac{A}{2 \kappa(h)} \bigg ( \bar{T}^{\psi}_{~ \psi} - \frac{1}{3} \bar{T}^{\tau}_{\tau} \bigg ),
\end{align}
where 
\begin{align}
\bar{T}^{\tau}_{~ \tau} = \frac{1}{2} \dot{h}^2 + 2 c_{dHdH} h^2 \dot{h}^2 - V(h) + 2 \bigg ( \ddot{f}(h) + \frac{3 \dot{A}}{A} \dot{f}(h) \bigg ) - 2 \ddot{f}(h), \\
\bar{T}^{\psi}_{~ \psi} = -\frac{1}{2} \dot{h}^2 - 2 c_{dHdH} h^2 \dot{h}^2  - V(h) + 2 \bigg ( \ddot{f}(h) + \frac{3 \dot{A}}{A} \dot{f}(h) \bigg ) - 2\frac{\dot{A}}{A} \dot{f}(h).
\end{align}
Using the above formulas and (\ref{bTupup}) we may write
\begin{align}
\label{eq_A_comp}
\ddot{A} = \frac{A}{2 \kappa(h)} \bigg [ - \frac{2}{3} \dot{h}^2 - \frac{5}{3} c_{dHdH} h^2 \dot{h}^2 - \frac{2}{3} V(h)
+ \frac{4}{3}  \bigg ( \ddot{f}(h) + \frac{3 \dot{A}}{A} \dot{f}(h) \bigg ) - 2\frac{\dot{A}}{A} \dot{f}(h)  
+ \frac{2}{3} \ddot{f}(h) 
\bigg ].
\end{align}
On the other hand, the equation of motion of the Higgs field is
\begin{align}
\label{eq_h}
\square h &+ 2 c_{dHdH} h \square h^2 - \frac{d V(h)}{dh} - R \frac{d f(h)}{dh} = 0, \\
\label{eq_h_comp}
\ddot{h} &+ \frac{3 \dot{A}}{A} \dot{h} + 2 c_{dHdH} h \bigg ( 2 h \ddot{h} + 2 \dot{h}^2 + \frac{3 \dot{A}}{A} 2 h \dot{h}
\bigg ) - m_{H}^2 h - 3a_3 h^2 - \frac{\lambda_{H}}{3!} h^3 - 6 c_{6} h^5 + \nonumber \\
&- R \bigg ( 2 \xi_{H} h + 4 c_{HH} h^3 \bigg ) = 0.
\end{align}
The system of equations (\ref{eq_A_comp}) and (\ref{eq_h_comp}) is the one that will be solved numerically to obtain profiles for the scale factor $A(\tau)$ and scalar field $h(\tau)$. To this end, we used the Maple software and its built-in ordinary differential equations (ODE) solvers \cite{Maple2020}. But before this can be done we should make the field $h$, the Euclidean time $\tau$ and various constants appearing in the action (\ref{cEFT_euc}) dimensionless. To this end, we use the following rescaling:
\begin{align}
\label{Mrescaling}
&\overline{M}^{-2} \overline{M}_{Pl}^2 \rightarrow \overline{M}_{Pl}^2, \quad \overline{M} \tau \rightarrow \tau, \quad \overline{M}^{-2} R \rightarrow R,   \quad \overline{M}^{-1} h \rightarrow h, \nonumber \\
& \quad  \overline{M}^{-1} m_H \rightarrow m_H, \quad
\overline{M}^{2} c_{dHdH} \rightarrow c_{dHdH}, \quad  \overline{M}^2 c_{HH} \rightarrow c_{HH},\nonumber \\
& \quad \overline{M}^{-1} a_3 \rightarrow a_3,  \quad \overline{M}^{2} c_{6} \rightarrow c_{6}, \quad \overline{M}^{-4} c_0 \rightarrow c_0,
\end{align} 
where we decided that for a better readability of the paper we will use the same symbols for both dimensionfull and dimensionless quantities. Since in our analysis we will always be using dimensionless quantities this should not lead to any confusion. 
In the above, $\overline{M}$ represents some scale of mass dimension one. Usually in the literature this is set to be the reduced 
Planck mass $\overline{M}_{Pl}$. Yet, in our case we have another energy/mass scale to consider, 
namely $m_{X}$. This stems from the fact that our effective theory is valid for energies and curvatures smaller than the heavy particle mass. For this reason it is convenient to set $\overline{M} = m_{X}$.  

After discussing the rescaling of the variables and parameters needed for numerical computations let us present the details of the numerical setup. 
Let us start by pointing out that basic equations that we intend to solve, namely (\ref{eq_h}) and an appropriate linear combination of equations
(\ref{eq_G00}) and (\ref{eq_G11}) that leads to (\ref{eq_A}) possess singularities at $\tau=0$ where we should 
have $A(\tau)_{| \tau =0} =0$. This implies the following set of boundary conditions:
\begin{align}
\label{bc_t0}
h(\tau)_{| \tau=0} = v_H, \qquad \dot{h}(\tau)_{| \tau =0} = 0, \nonumber \\
A(\tau)_{| \tau = 0} = 0, \qquad \dot{A}(\tau)_{| \tau =0} = 1,
\end{align}
where $v_H$ is a constant. The first set of the above conditions implies that the scalar field is constant, but $v_{H}$ does not need to be 
the minimum of the potential, in fact it is not. The second set translates to the requirement that there is no conical singularity at $\tau =0$
and the Ricci scalar as given by (\ref{Ricci_sc}) is finite (this is ensured by $\dot{A}(\tau)_{| \tau =0} = 1$).
(\ref{bc_t0}) contains four conditions which is precisely the number of conditions we can specify to solve the system of equations 
(\ref{eq_A_comp})--(\ref{eq_h_comp}). Nevertheless, we are looking for a specific form of the solution to the scalar field equation, namely
the bounce solution. It is characterized by two conditions $h(\tau)_{| \tau=0} = v_H$ and $h(\tau)_{| \tau \rightarrow \infty} = v_{\infty}$,
where $v_{\infty}$ is some constant, in our case $v_{\infty}=0$. To obtain such a solution we treated $v_H$ as a free parameter.
We chose some value for it and solved our system of equations. Then we repeated it until we found $v_{H}$ for which $h$ goes to zero at 
large $\tau$. This procedure of solving a system of ODEs is often called 'shooting' and its description can be found for example in 
\cite{numerical_recip}. As an additional point let us note that if $h(\tau)$ goes sufficiently fast to zero, and provided that $c_0 =0$, the spacetime becomes asymptotically flat. 

What we described above is the general algorithm for solving our problem. Let us now focus on some details of its implementation.
Firstly, it is not feasible to start numerical integration from the singular point therefore we will start not at $\tau=0$ but at $\tau =\epsilon$,
where $\epsilon = 10^{-5}$. To obtain the boundary condition at this point we look for the series solution to 
(\ref{eq_A_comp})--(\ref{eq_h_comp}) at $\tau=\epsilon$. To this end, we use the following ansatz for $h$ and $A$:
\begin{align}
\label{hseries}
 h(\tau) = v_H + \alpha_2 \tau^2 + \alpha_3 \tau^3 + \alpha_4 \tau^4, \\
\label{Aseries}
A(\tau) \equiv \tau + Ap(\tau), \qquad Ap(\tau) = \beta_3 \tau^3 + \beta_4 \tau^4.
\end{align}
The form of the series was chosen in such a way that $h$ and $A$ fulfill boundary conditions (\ref{bc_t0}).
Moreover, the splitting of $A$ was dictated by the need to isolate the asymptotic behavior for $\tau \rightarrow \infty$,
namely $A(\tau)_{| \tau \rightarrow \infty} \sim \tau$. This form turns out to be more convenient for numerical calculations.
Having put the above ansatz into the equations we found solutions that satisfy them up to terms linear in $\tau$. This implies that 
we introduce an error of the order of $\mathcal{O}(\epsilon^2)$ to our numerical scheme regardless of the accuracy of numerical integration. 
Then, as our starting condition for integration we set $h(\epsilon)$ and $Ap(\epsilon)$, let us note that the value of $h(\epsilon)$
still depends on $v_H$. Secondly, we cannot integrate up to $\tau = \infty$, therefore need some stopping criteria. The obvious one is $h(\tau)_{| \tau_{max}} =0$ but this is insufficient. As a supplementary criterion we choose $\dot{h}(\tau)_{| \tau_{max}} < \epsilon_{\tau}$, where $\epsilon_{\tau}$ is a small parameter. We tested our code for a value of this parameter from within the range $\langle 10^{-6}, 10^{-8}\rangle$ and found no significant difference in the obtained results. For this reason we chose $\epsilon_{\tau} = 10^{-6}$. This set of conditions implies that after some Euclidean time $h$ will decrease rapidly and then approach $h(\tau) = 0$ in a very gentle way, as so after reaching $\tau_{max}$ it can be glued to
the analytic solution $h(\tau) = 0$. To sum up, the described procedure gives us the value of $\tau_{max}$ up to which we need to 
perform the integration. As far as $Ap(\tau)$ is concerned, it turns out that the abovementioned conditions lead to satisfactory behavior of this function.

\section{Results}
\label{sec:results}

After explaining our analytical and numerical setups in the previous section let us focus on the obtained results.
As was mentioned above, we are looking for solutions of equations of motion that represent bounce (or soliton-like) behavior
for the scalar field and the corresponding metric function. From the physical standpoint this type of solutions represents a field that
is close to true vacuum at $\tau=0$ and then decays to false vacuum at large $\tau$. One of the standard ways of obtaining the vacuum decay probability (in the context of the Euclidean approach used in this study) is to calculate the action on the solutions 
of equations of motion and then the vacuum decay probability (per unit volume) is given by
\begin{align}
\label{G_decay}
\Gamma = \mathcal{A} e^{-\mathcal{S}}.
\end{align}
In the above formula $\mathcal{A}$ represents a prefactor coming from a quantum correction to the action. 
In flat spacetime it may be either accounted for by dimensional arguments or calculated numerically.
Let us also note that there is possibility that it could be calculated analytically with the help of the heat kernel method,
but this is the problem for a separate study. As for the calculation of the prefactor in curved spacetime, there is an
additional difficulty in accounting contributions of the fluctuation of the metric field.  
Returning to the case at hand, since the prefactor represents a higher order correction to $e^{-\mathcal{S}}$ 
we will not discuss it any further. 
Let us now focus on the leading contribution to the vacuum decay that comes from the exponential factor. In this factor $\mathcal{S}$
represents a difference between the Euclidean action calculated at the bounce solution and the same action calculated at the solution representing 
false vacuum $\mathcal{S} = S_{cEFT}^{euc}(bounce) - S_{cEFT}^{euc}(false~vacuum)$, where $S_{cEFT}^{euc}$ is 
given by (\ref{cEFT_euc}). At this point let us remark that in our study we will set the $c_0$ parameter in the potential of the scalar field to be equal to $0$, what is equivalent to setting the cosmological constant to zero. 
Moreover, for our model the false vacuum solution is given by $h(\tau) = 0$,
therefore it implies (with $c_0 = \Lambda = 0$) that $R = 0$. This means that $S_{cEFT}^{euc}(false~vacuum) = 0$. This also implies the integration in $S_{cEFT}^{euc}(bounce)$ over a finite interval $\tau~\epsilon~\langle0,\tau_{max}\rangle$. This is so because for $\tau$
larger than $\tau_{max}$ we have by definition $h(\tau)_{|\tau > \tau_{max}} = 0$ and therefore $R(\tau)_{|\tau > \tau_{max}} = 0$.

As far as the boundary terms are concerned, we should supplement the action given by (\ref{cEFT_euc}) with the Gibbons-Hawking-York
boundary term \cite{PhysRevLett.28.1082,PhysRevD.15.2752,BARVINSKY1996305}
\begin{align} 
S_{GHY} = 2 \int d^3 x \sqrt{\bf{h}} \kappa(h) K,
\end{align}
where $\kappa(h)$ was defined in (\ref{kappah}), $\bf{h}$ is the induced metric on the boundary hypersurface and $K$ is the trace 
of its extrinsic curvature. For our action and the metric ansatz it may be explicitly written as
\begin{align}
S_{GHY} = - 6 \pi^2 \kappa(h) \left(A^2 \ddot{A}\right) \Big\rvert_{\tau=0}^{\tau =\infty}.
\end{align} 
In the formula above we integrated over the $S^3$ coordinates. Since for our setup the fields and metric configurations for both the false vacuum 
and the bounce configuration are identical outside $\tau = \tau_{max}$, there is no contribution to the exponent in (\ref{G_decay})
from the hypersurface $\tau = \infty$.  As far as $\tau=0$ hypersurface is concerned, $S_{GHY}$ vanishes there due to the vanishing of the metric function $A$, therefore this also does not contribute to $\Gamma$.
Before we go into the details of the obtained results let us mention the problem of obtaining the Lorentzian counterpart of the 
metric and scalar field profiles. First of all let us stress that we regard the Euclidean approach as a convenient tool to calculate 
the decay rate $\Gamma$. Therefore, although the real time evolution of the nucleated true vacuum bubble is an important topic, it goes 
beyond the scope of this paper.  Nevertheless,  let us briefly mention that to obtain the Lorentzian counterpart we may follow 
 \cite{PhysRevD.21.3305} and \cite{HAWKING199825,Burda:2016mou}. Especially the two latter articles turn out to be very 
 useful in this regard. 

\subsection{Description in language of the cEFT coefficients}

\subsubsection*{Influence of the $\xi_H$ on the vacuum stability}

\begin{figure}[h]
\centering
\subfloat[]{
  \includegraphics[width=.9\textwidth]{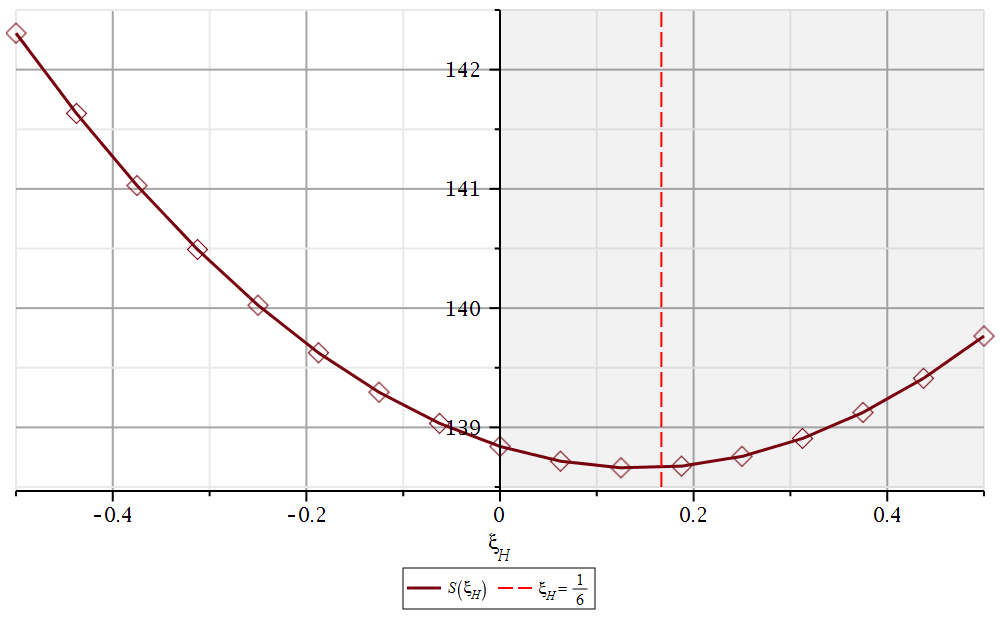}
\label{fig1a}
}
\quad
\subfloat[]{
  \includegraphics[width=.9\textwidth]{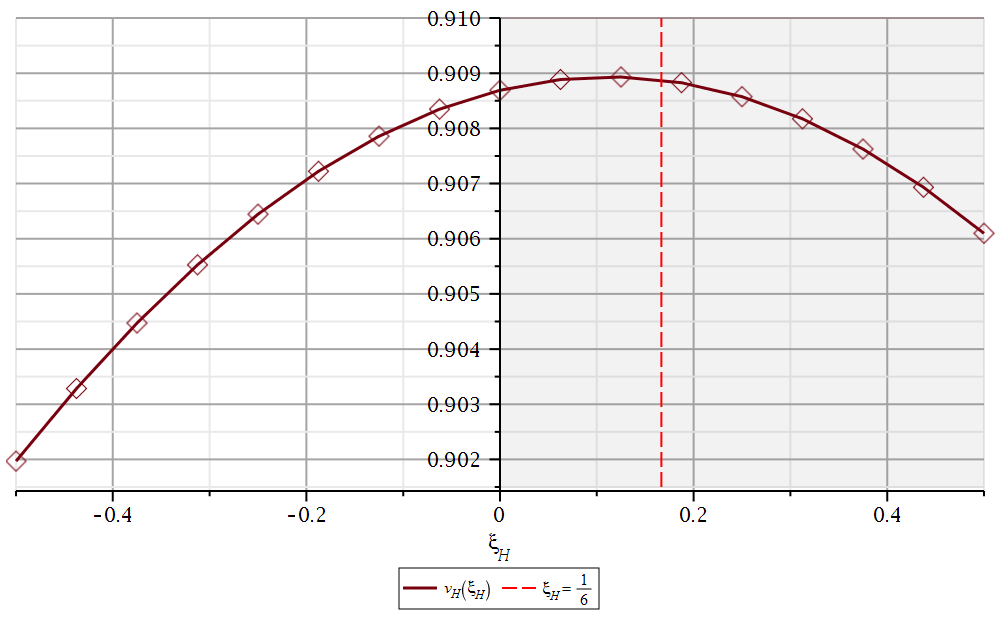}
\label{fig1b}
}
\caption{(a) The action calculated on solutions of EOM and (b) the value of the $h(\tau)$ field at $\tau =0$
as functions of $\xi_{H}$. The remaining parameters are $\bar{M}_{Pl}^2 = 10^2, 
\lambda_{H} = 6.0, m_{H}^2 = 0.2, a_{3} = -0.4, c_{HH} = 0, c_{6} = c_{dHdH} = 0.0, c_{0} = 0.0$.}
\label{fig1}
\end{figure}

Below we present and discuss the results of our calculations. 
Figure~\ref{fig1a} presents the influence of the non-minimal coupling term $\xi_H h^2 R$ on vacuum stability.
We restricted ourselves to the case $\xi_H \leqslant 0$ (and $c_{HH} \leqslant 0$), since for this $\kappa(h)$ as defined by (\ref{kappah}) is always positive.
If we want to interpret $\kappa(h)$ as the effective Planck mass this restriction means that we demand that it is always positive.  
From this figure we infer that as $|\xi_H|$ increases $\mathcal{S}$ increases, hence large negative $\xi_H$ stabilizes false vacuum.
This is in agreement with the results obtained in the work \cite{PhysRevD.95.025008} (discussed also in the review paper 
\cite{10.3389/fspas.2018.00040}). Moreover, for completeness, we presented also results for $\xi_H >0$ (the shaded region
in Figures \ref{fig1a} and \ref{fig1b}). From it we see that the minimum of $\mathcal{S}$ lays slightly below $\xi_H = \frac{1}{6}$,
which is represented by a red vertical line.
On the other hand, from Figure \ref{fig1b} we see that in the $\xi_H \leqslant 0$ region an increase of $|\xi_H|$ leads 
to the decrease of the $v_H$ parameter (this parameter represents the value of the scalar field at $\tau=0$). 
From the shaded portion of the aforementioned plot we see that the maximum of $v_{H}$ also lays slightly below 
$\xi_H = \frac{1}{6}$.

\begin{figure}[h]
\centering
\subfloat[]{
  \includegraphics[width=.8\textwidth]{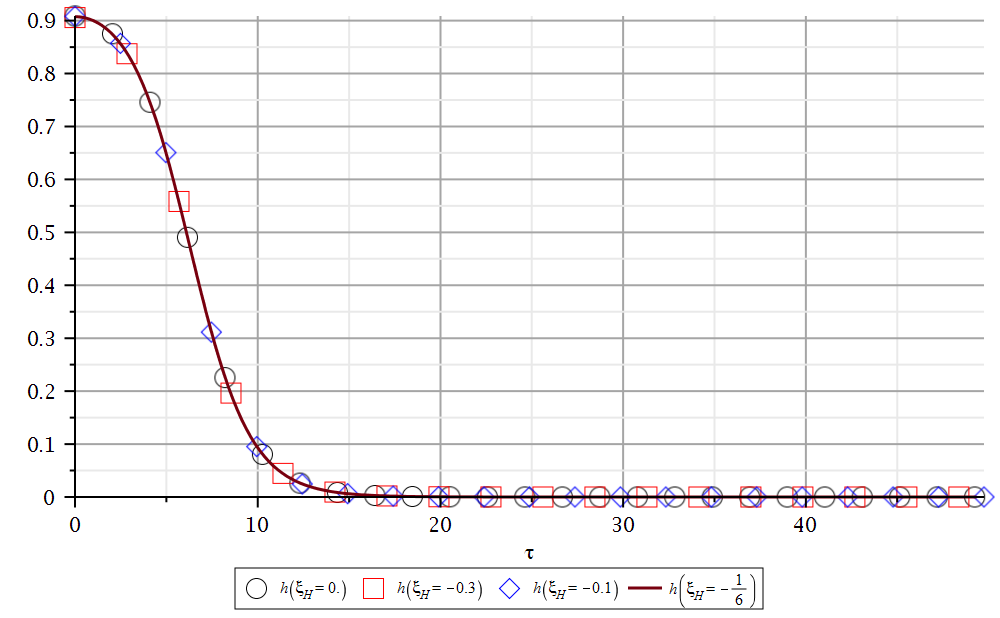}
\label{fig2a}
}
\quad
\subfloat[]{
  \includegraphics[width=.8\textwidth]{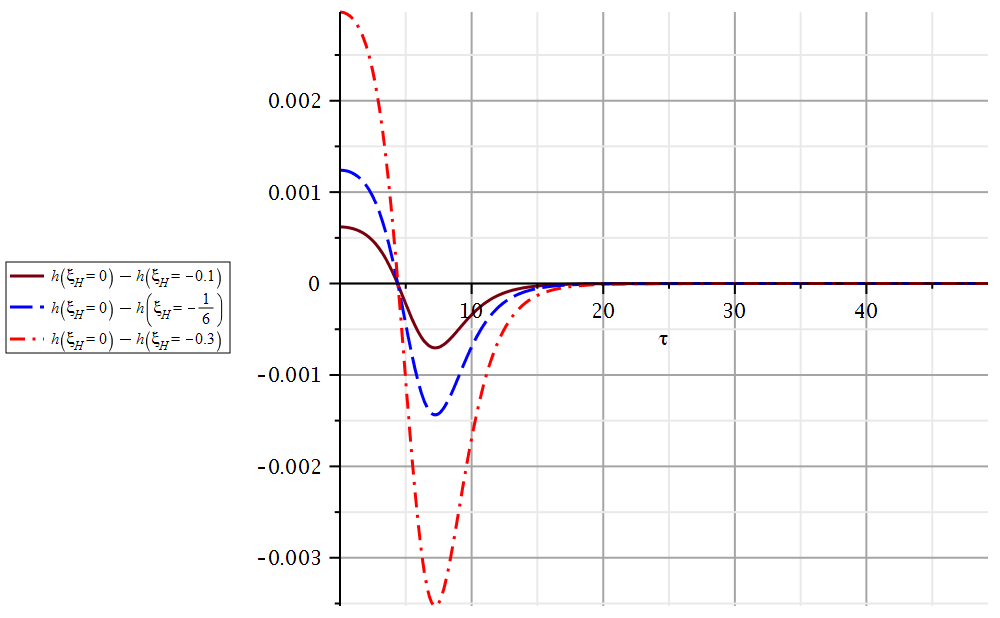}
\label{fig2b}
}
\caption{(a) Profiles of the scalar field $h(\tau)$ for various $\xi_H$ and (b) differences between the profiles for $\xi_H = 0$
and $\xi_H \neq 0$. The remaining parameters are $\bar{M}_{Pl}^2 = 10^2, 
\lambda_{H} = 6.0, m_{H}^2 = 0.2, a_{3} = -0.4, c_{HH} = 0, c_{6} = c_{dHdH} = 0.0, c_{0} = 0.0$.}
\label{fig2}
\end{figure}

In Figure~\ref{fig2a} we plotted the profile of the scalar field $h$ for various values of the non-minimal coupling parameter.
Differences in the profiles are very small and for large $|\xi_H|$ they are the biggest in two regions. One is at $\tau =0$ 
and the second one is where the field changes its value most rapidly. 
Moreover, comparing Figures~\ref{fig2a} and \ref{fig2b} we may conclude that the relative change is bigger in the 
region in which scalar field varies rapidly. For example, for $\xi_H = -0.3$ we may roughly estimate 
$\Delta h = \frac{|h(\tau \approx 7, \xi_H =0) - |h(\tau \approx 7, \xi_H =- 0.3)|}{|h(\tau \approx 7, \xi_H =0)} 
= \frac{0.0035}{0.3} \approx 0.012$, while for $\tau=0$ similar calculation gives us $\Delta h \approx 0.003$.

\begin{figure}[h]
\centering
\subfloat[]{
  \includegraphics[width=.8\textwidth]{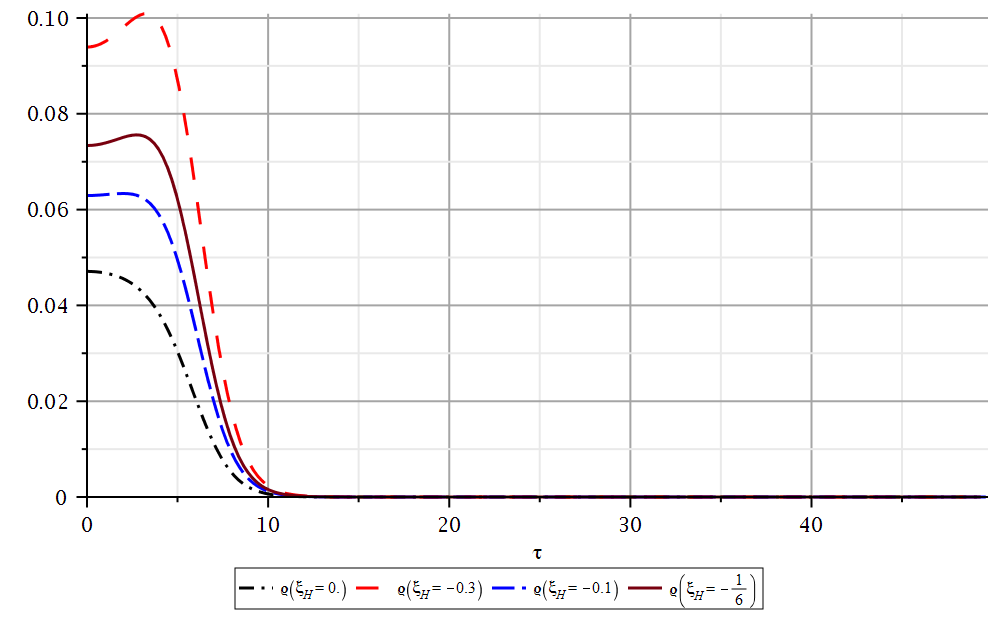}
\label{fig3a}
}
\quad
\subfloat[]{
  \includegraphics[width=.8\textwidth]{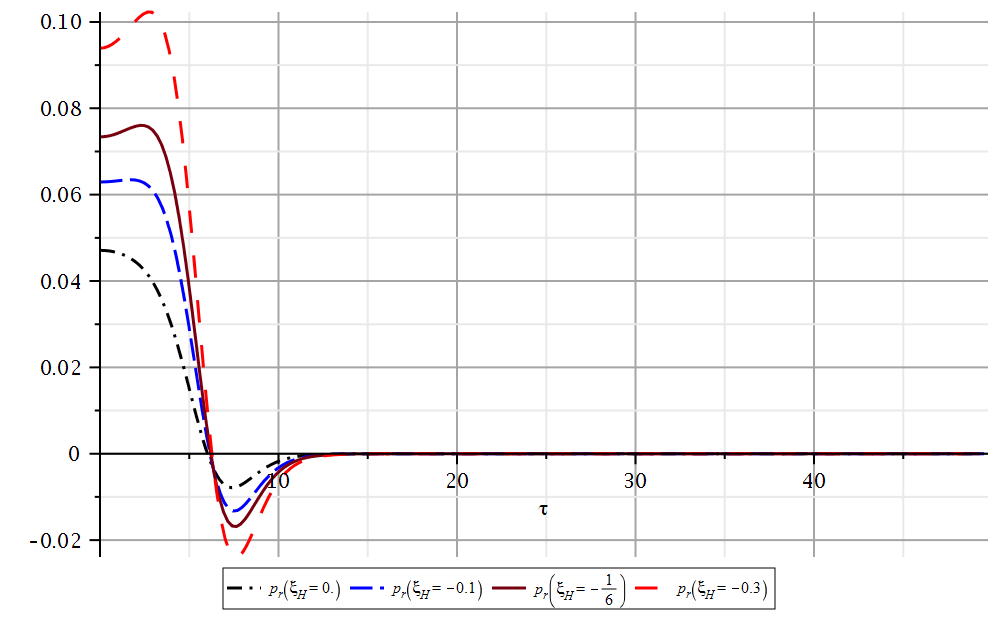}
\label{fig3b}
}
\caption{(a) Energy densities defined as $\rho \equiv \frac{\overline{M}_{Pl}^2}{\kappa(h)} \overline{T}^{\tau}_{~~ \tau}$ 
for various $\xi_{H}$. The remaining parameters are 
$\overline{M}_{Pl}^2 = 10^2, \lambda_{H} = 6.0, m_{H}^2 = 0.2, 
a_{3} = -0.4, c_{HH} = 0, c_{6} = c_{dHdH} = 0.0, c_{0} = 0.0$. 
(b) The pressure defined as $p_{r} \equiv  \frac{\overline{M}_{Pl}^2}{\kappa(h)} \overline{T}^{\psi }_{~~\psi}$,
 parameters are the same like for $\rho$.}
\label{fig3}
\end{figure}

In Figure~\ref{fig3a} we plotted the change in one of the observables mentioned in the Introduction, namely the energy density 
defined as $\rho \equiv \frac{\overline{M}_{Pl}^2}{\kappa(h)}\bar{T}^{\tau}_{~~\tau}$, 
where  $\bar{T}^{\tau}_{~~\tau}$ is the appropriate component 
of the energy-momentum tensor as defined in (\ref{bTupup}). From this plot we 
may see that for a minimally coupled scalar field $\rho$ is always positive and its maximum value is at $\tau =0$. As we increase $\xi_H$
this maximum value starts to increase and for sufficiently large negative $\xi_H$ it is shifted to $\tau \neq 0$.
Meanwhile, for $\tau \rightarrow \tau_{max}$ $\rho$ approaches zero which is in agreement with our boundary 
condition $h(\tau)_{|\tau=\tau_{max}} =0$.
This together with the fact that the overall shape of the profiles of the scalar field does not differ significantly for various $\xi_H$
leads us to the conclusion that the dominant contribution to the action associated with the non-minimal coupling is through 
the derivative term that contributes to the energy-momentum tensor and therefore to the geometry itself. 

\begin{figure}[h]
\centering
\subfloat[]{
  \includegraphics[width=.8\textwidth]{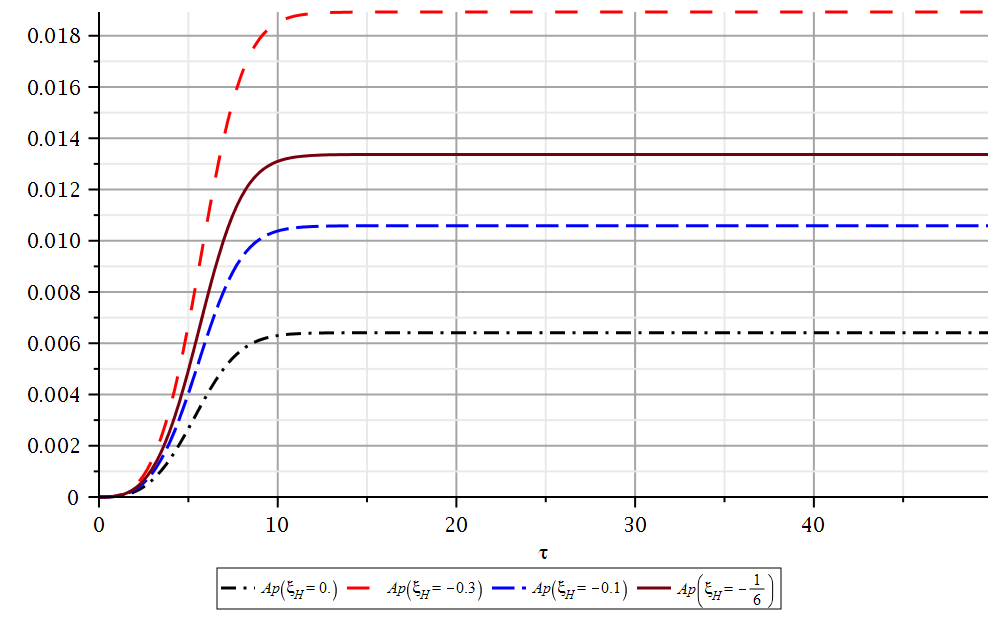}
\label{fig4a}
}
\quad
\subfloat[]{
  \includegraphics[width=.8\textwidth]{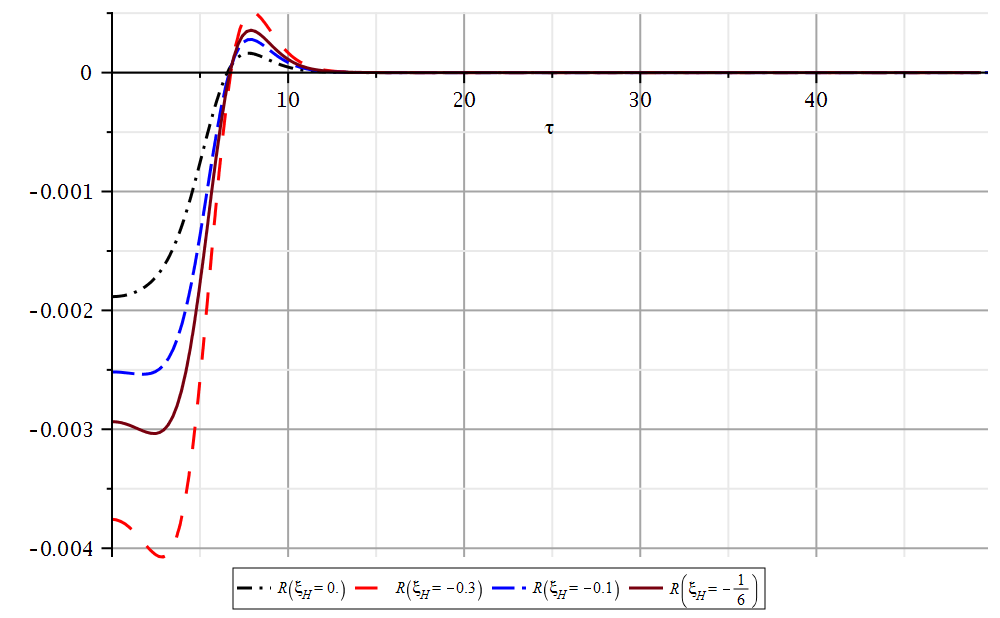}
\label{fig4b}
}
\caption{(a) Profiles of the metric function $A(\tau) \equiv \tau + Ap(\tau)$ for various $\xi_H$ and (b) the Ricci scalar $R$ for various $\xi_H$. The remaining parameters are 
$\bar{M}_{Pl}^2 = 10^2, \lambda_{H} = 6.0, m_{H}^2 = 0.2, a_{3} = -0.4, c_{HH} = 0, c_{6} = c_{dHdH} = 0.0, c_{0} = 0.0$.}
\label{fig4}
\end{figure}

This is also visible in Figures~\ref{fig4a} and \ref{fig4b}. In Figure~\ref{fig4b} we plotted the second observable, namely 
the Ricci scalar for various $\xi_H$. 
We may infer from it that for the minimally coupled scalar $R$ is negative in the vicinity of $\tau=0$, then it has a local maximum
at around $\tau \approx 6.0$ (this roughly corresponds to the bubble width at half maximum as can be seen from Figure~\ref{fig2a}), and then it goes to zero as expected due to the boundary conditions. 
Meanwhile, for large negative $\xi_H$ the Ricci scalar develops its minimum away from $\tau=0$
and then attains a local maximum and eventually decreases to zero. 
In our opinion this cannot be explained solely by the small difference in the value of the field at $\tau=0$ which will contribute to the 
equations of motion and the action through the potential term. Therefore, we conclude (as was stated before) that the actual effect 
must be due to the gradients of the field that influence geometry.
Another interesting fact that can be inferred from Figure~\ref{fig4a} is that the Ricci scalar is not a monotonic function of the
Euclidean time. Moreover, from this figure we may also see that $R$ changes its sign close to the bubble wall.
From Figure~\ref{fig3b} we may see that the region of positive $R$ is the same as the region in which pressure 
defined as $p_{r} \equiv \frac{\overline{M}_{Pl}^2}{\kappa(h)}\bar{T}^{\psi}_{~~\psi}$ attains a negative value.

In Figure~\ref{fig4a} we plotted the profile of the metric function $Ap(\tau)$. Again we see that for the minimally coupled 
scalar the $Ap(\tau)$ functions tend to a positive constant and for large negative $\xi_H$ this asymptotic value is bigger.
As a side note, let us point that since the metric function $A(\tau)$ grows up as $\tau + Ap(\tau)$, the obtained spacetime
is an asymptotically flat one.

\begin{figure}[h]
\centering
\subfloat[]{
  \includegraphics[width=.9\textwidth]{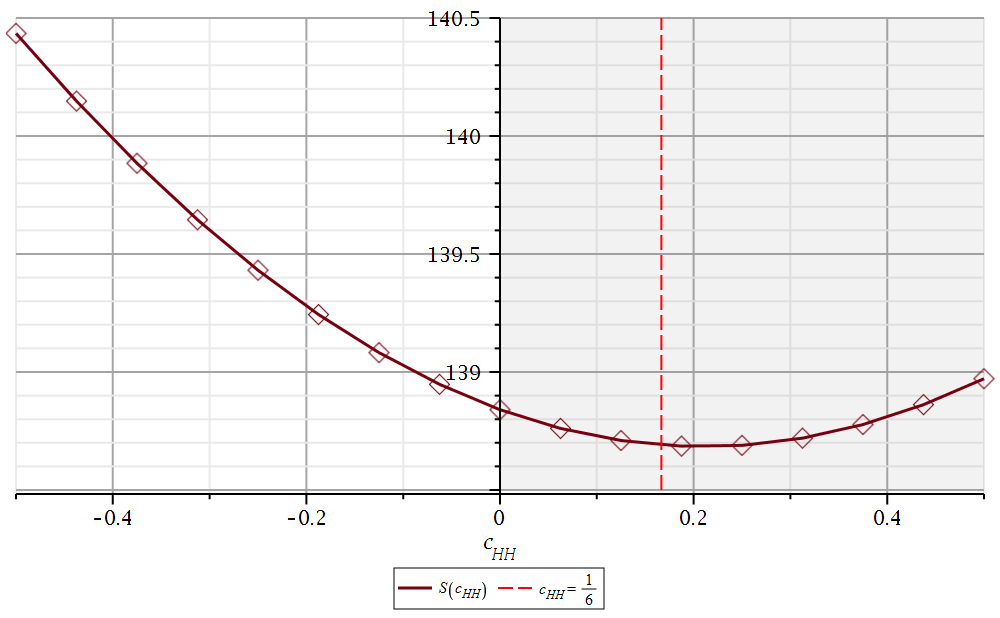}
\label{fig5a}
}
\quad
\subfloat[]{
  \includegraphics[width=.9\textwidth]{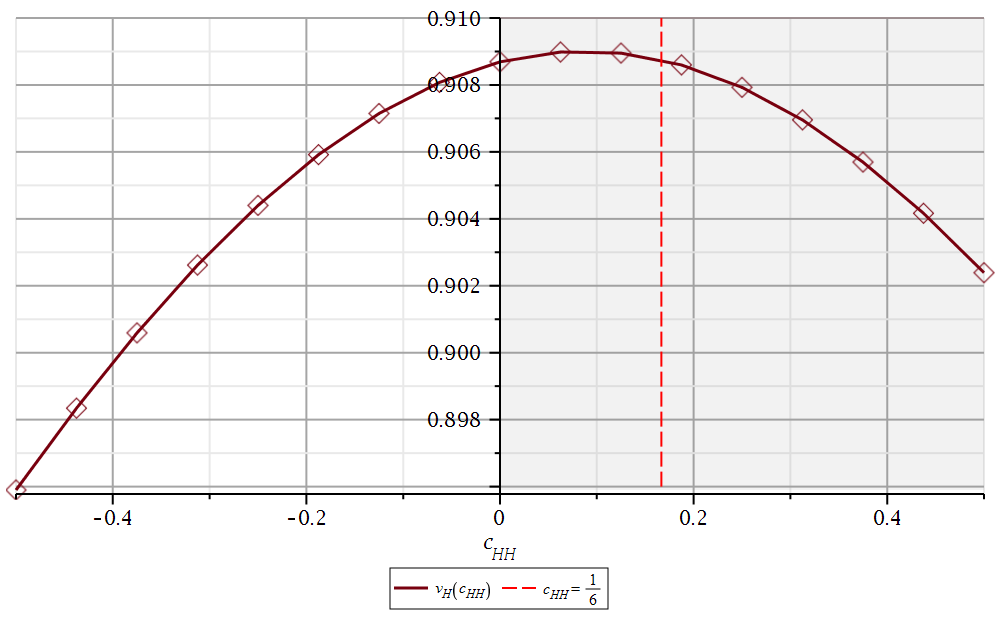}
\label{fig5b}
}
\caption{(a) The action calculated on solutions of EOM and (b) the $h(\tau)$ field at $\tau =0$ as functions of $c_{HH}$. 
The remaining parameters are $\bar{M}_{Pl}^2 = 10^2, 
\lambda_{H} = 6.0, m_{H}^2 = 0.2, a_{3} = -0.4, \xi_{H} = 0, c_{6} = c_{dHdH} = 0.0, c_{0} = 0.0$.}
\label{fig5}
\end{figure}

After discussing the results presented in Figures~\ref{fig1} to \ref{fig5} we may summarize our findings in the following conclusion. 
The stabilization of the false vacuum for large values of $|\xi_H|$ may be connected to the fact that the 
energy density inside the true vacuum bubble also increases with $|\xi_H|$, therefore it is more costly to create such a bubble
than for the $\xi_H =0$ case. Moreover, the scalar field gradient close to the bubble wall enhances the maximum value 
attained by the Ricci scalar at this location. We connected this region of positive $R$ to the region in which the 
pressure associated  with the scalar field is negative. As a side note, let us point out that this also indicates 
that the thin-wall approximation may be inapplicable in this case, as was pointed out in \cite{PhysRevD.95.025008}.

Let us at this point pause and tackle one more technical problem. We work within the framework of the effective field theory 
in curved spacetime.
This implies that there is an energy scale $\mu_{>}$ above which our theory is no longer valid. 
For our case this scale is proportional to the mass of the 
integrated out heavy particle $X$. Due to the employed rescaling (\ref{Mrescaling}) we may set the numerical value of $\mu_{>}$ to be equal to $1$. Translating this energy scale to the energy density $\rho_{>} \sim \mu_{>}^4$ we need the following constraints to be fulfilled: $\frac{\rho}{ \rho_{>}}<1$ and $\frac{R}{\mu_{>}^2} < 1$. As we may see from Figures~\ref{fig3a} and \ref{fig4b}, despite the fact that the second minimum of our potential is roughly at $h \sim 1 \sim \mu_{>}$ the abovementioned constraints are fulfilled.  
As for the higher order operators not accounted for by the action (\ref{action_cEFT_h}), they are suppressed by the higher powers of $m_{X}^2$. The only subtlety concerns the gravity mediated operators containing higher powers of curvature scalars.
Yet, as was mentioned above, the smallness of the Ricci scalar and energy density indicate that they also will be sufficiently suppressed.  

\subsubsection*{Influence of the $c_{HH}$ on the vacuum stability}

Now we will focus on the influence of the $c_{HH}R h^4$ term on the vacuum stability. From Figure~\ref{fig5a}
we may see that the influence of $c_{HH}$ on the false vacuum stability is similar to the $\xi_H$ case.
The overall improvement in the stability is smaller than for $\xi_H$ and from analysis of the shaded region corresponding
to the $c_{HH}>0$ case we infer that the minimum of $\mathcal{S}$ lays above $c_{HH} = \frac{1}{6}$ value.
Meanwhile, maximum $v_H$ is attained for $c_{HH} < \frac{1}{6}$ which is visible in Figure~\ref{fig5b}.
Comparing data from Figures~\ref{fig9a} and \ref{fig9b} we may obtain relative differences between 
scalar field profiles at the point of a biggest difference among them. 
For the case of $c_{HH} = -0.3$ they are 
$\Delta h_{|\tau=0} =  \frac{|h(\tau=0,c_{HH}=0) - h(\tau=0,c_{HH}=-0.3)|}{h(\tau=0,c_{HH}=0)}
\approx \frac{0.006}{0.9} \approx 0.007$   and for $\tau\approx 6$ we have 
$\Delta h_{|\tau \approx 6} \approx \frac{0.002}{0.3} \approx 0.007 $. This is in contrast with the $\xi_H$ case
where the relative difference in profiles close to the bubble wall was bigger than at $\tau =0$.
This implies that for the $c_{HH}$ case the overall difference in the profiles $\Delta h$ is spread more uniformly
inside the bubble.

An analysis of Figures~\ref{fig8a} and \ref{fig8b} follows like in the $\xi_H$ case. Again we see that the
Ricci scalar develops a local minimum away from $\tau =0$ as we increase $|c_{HH}|$ and it reaches
a positive maximum value at a rough location of the bubble wall. The same region is also the locus of 
negative pressure as can be seen from Figure~\ref{fig7b}.

Turning to Figure~\ref{fig7a} we see that an increase in the stability of the false vacuum may be tracked to the larger 
energy density inside the true vacuum bubble and therefore larger energetic cost of creating such a bubble.  

\begin{figure}[h]
\includegraphics[width=.9\textwidth]{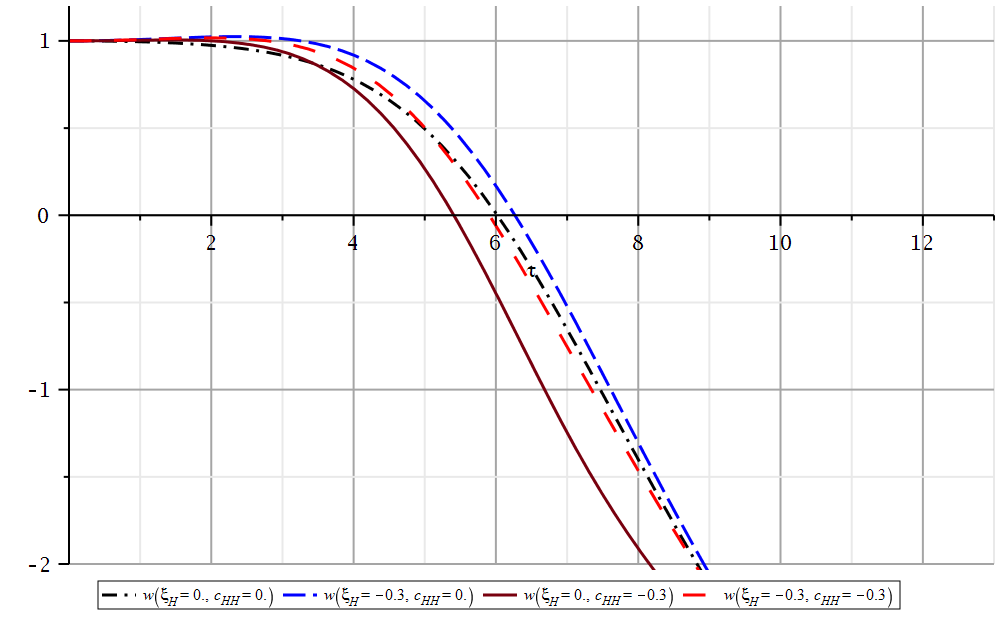}
\caption{Pressure to energy density rations $w = \frac{p_r}{\rho}$ for various ($\xi_H$ , $c_{HH}$ configurations.
The remaining parameters are $\bar{M}_{Pl}^2 = 10^2, \lambda_{H} = 6.0, m_{H}^2 = 0.2, a_{3} = -0.4, c_{6} = c_{dHdH} = 0.0, c_{0} = 0.0$.}
\label{fig6}
\end{figure}

At this point it is interesting to illustrate the difference between configurations with non-zero $\xi_H$ and $c_{HH}$.
To this end, we may look at Figure~\ref{fig6} where we plotted the pressure to energy density ratios 
defined as $w = \frac{p_r}{\rho}$ for the interior of the true vacuum bubble. From it we see that for 
$(\xi_H \neq 0, c_{HH}=0)$ configuration pressure decays slower than for the minimal case of $(\xi_H=0,c_{HH}=0)$.
On the other hand, for the $(\xi_H = 0, c_{HH} \neq 0)$ case the opposite is true. Interestingly, for a configuration 
with equal $\xi_H$ and $c_{HH}$ the distribution of the scalar field energy content among energy density and pressure
is most similar to the minimal case.


\begin{figure}[h]
\centering
\subfloat[]{
  \includegraphics[width=.8\textwidth]{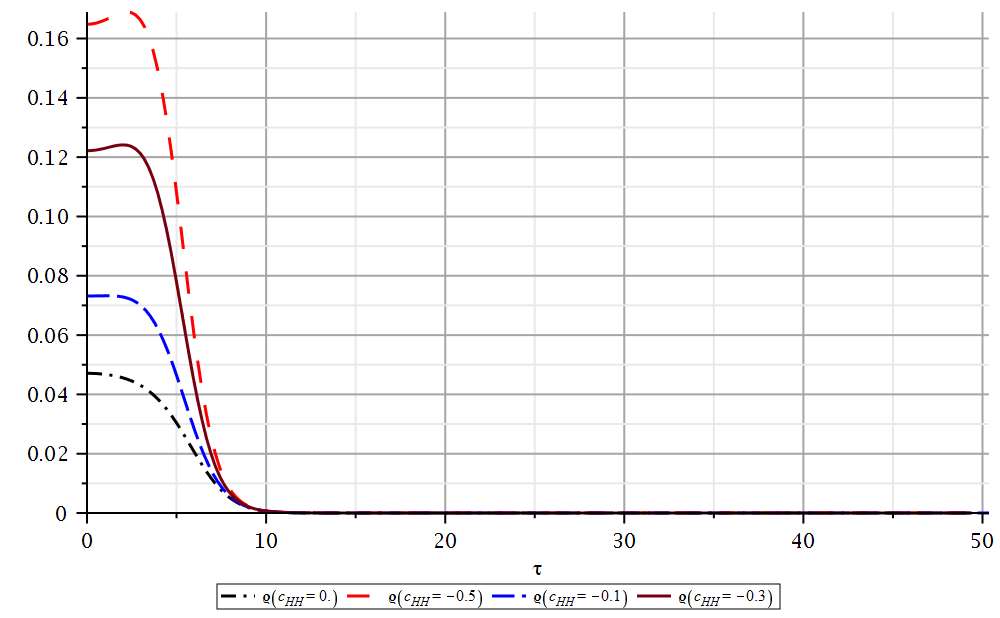}
\label{fig7a}
}
\quad
\subfloat[]{
  \includegraphics[width=.8\textwidth]{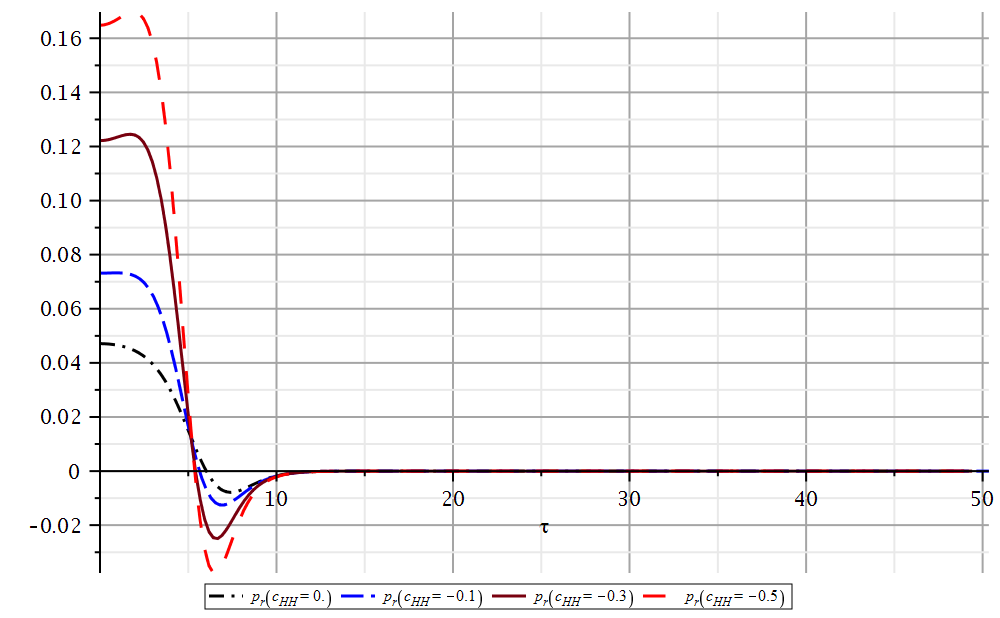}
\label{fig7b}
}
\caption{(a) Energy densities defined as $\rho \equiv \frac{\overline{M}_{Pl}^2}{\kappa(h)} \overline{T}^{\tau}_{~~ \tau}$ 
for various $c_{HH}$. The remaining parameters are $\bar{M}_{Pl}^2 = 10^2, \lambda_{H} = 6.0, m_{H}^2 = 0.2, a_{3} = -0.4, \xi_{H} = 0, c_{6} = c_{dHdH} = 0.0, c_{0} = 0.0$. 
(b) The pressure defined as $p_{r} \equiv  \frac{\overline{M}_{Pl}^2}{\kappa(h)} \overline{T}^{\psi }_{~~\psi}$,
 parameters are the same like for $\rho$.}
\label{fig7}
\end{figure}

\begin{figure}[h]
\centering
\subfloat[]{
  \includegraphics[width=.8\textwidth]{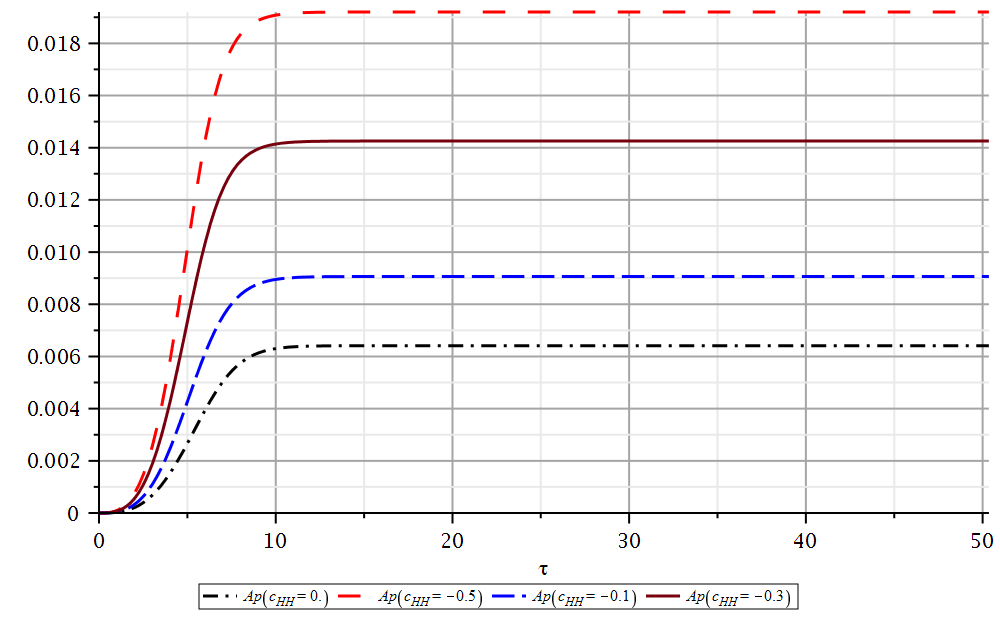}
\label{fig8a}
}
\quad
\subfloat[]{
  \includegraphics[width=.8\textwidth]{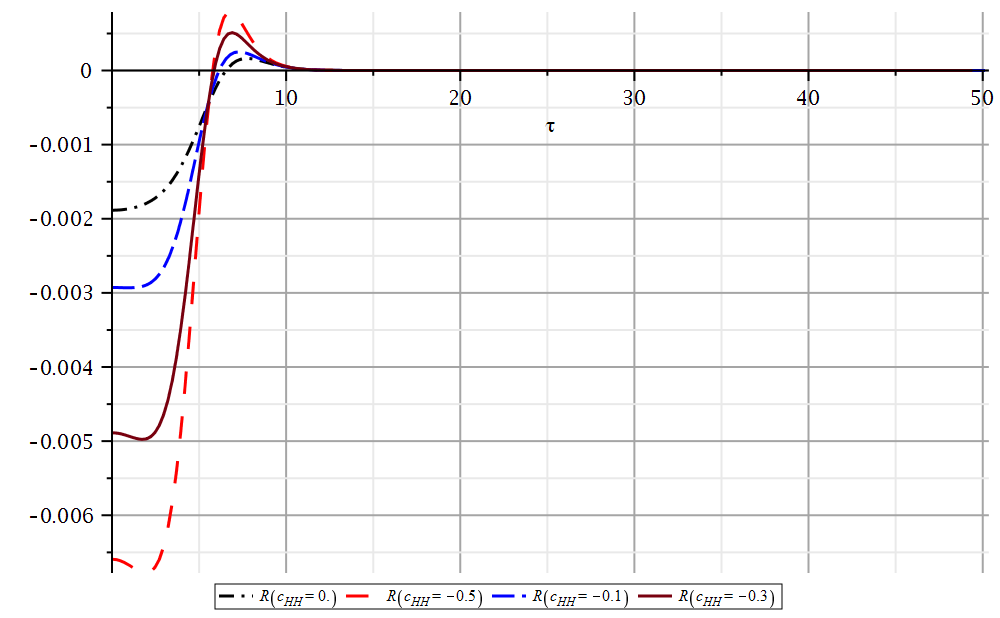}
\label{fig8b}
}
\caption{(a) Profiles of the metric function $A(\tau) \equiv \tau + Ap(\tau)$ and (b) the Ricci scalar $R$ for various $c_{HH}$. The remaining parameters are $\bar{M}_{Pl}^2 = 10^2, 
\lambda_{H} = 6.0, m_{H}^2 = 0.2, a_{3} = -0.4, \xi_{H} = 0, c_{6} = c_{dHdH} = 0.0, c_{0} = 0.0$.}
\label{fig8}
\end{figure}

\begin{figure}[h]
\centering
\subfloat[]{
  \includegraphics[width=.8\textwidth]{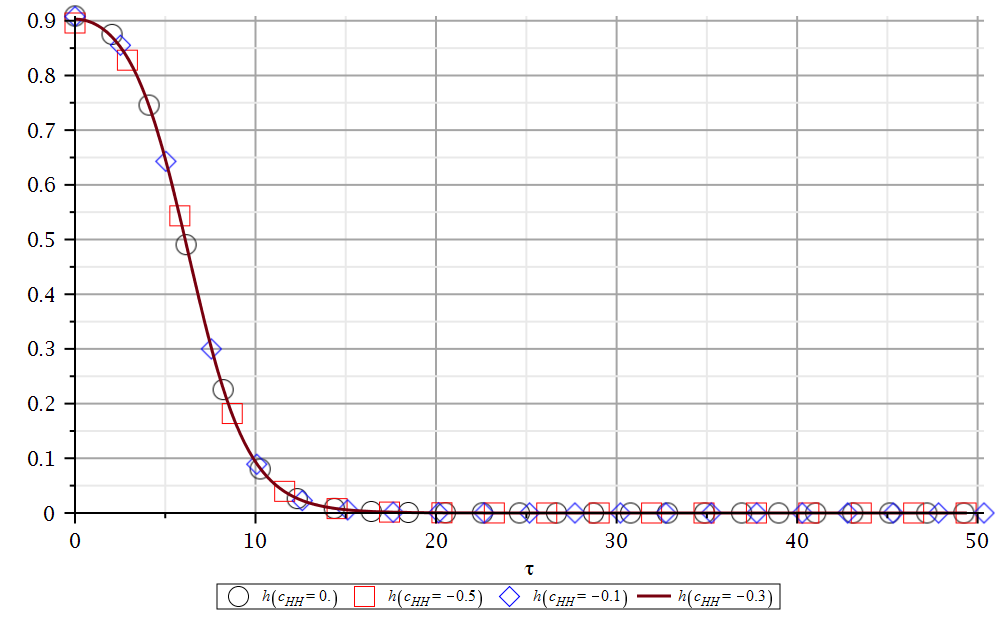}
\label{fig9a}
}
\quad
\subfloat[]{
  \includegraphics[width=.8\textwidth]{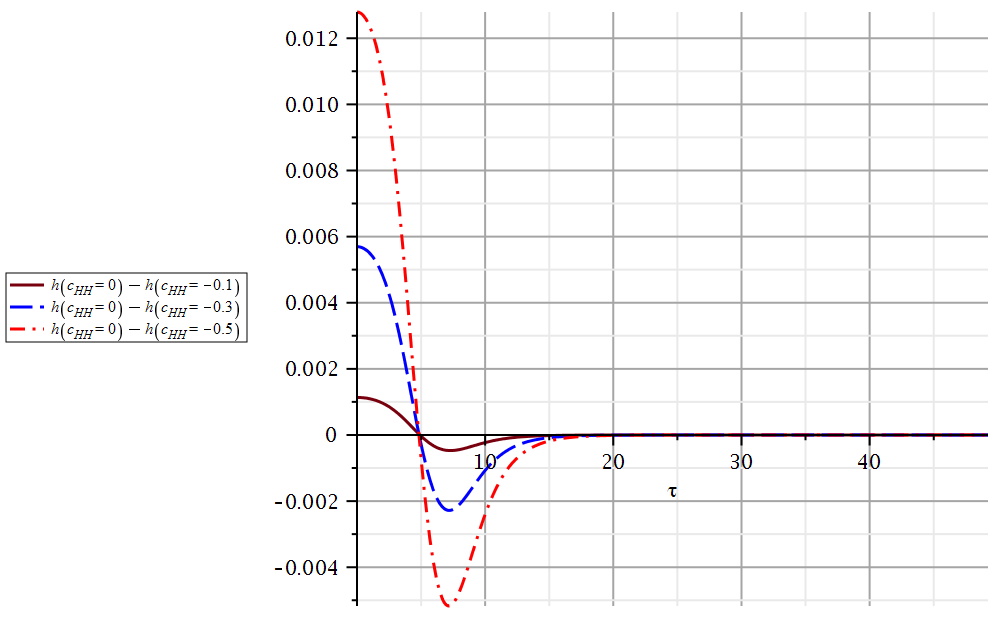}
\label{fig9b}
}
\caption{(a) Profiles of the scalar field $h(\tau)$ and (b) differences in profiles for various $c_{HH}$.
The remaining parameters are $\bar{M}_{Pl}^2 = 10^2, 
\lambda_{H} = 6.0, m_{H}^2 = 0.2, a_{3} = -0.4, \xi_{H} = 0, c_{6} = c_{dHdH} = 0.0, c_{0} = 0.0$.}
\label{fig9}
\end{figure}

\subsubsection*{Influence of the $c_6$ and $c_{dHdH}$ on the vacuum stability}

\begin{figure}[h]
\centering
\subfloat[]{
  \includegraphics[width=.8\textwidth]{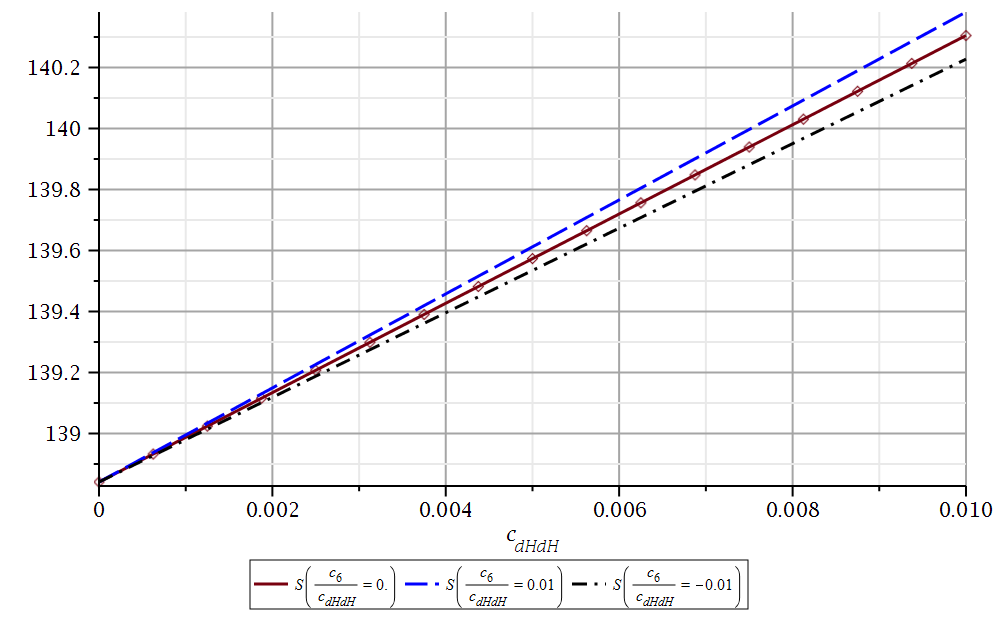}
\label{fig10a}
}
\quad
\subfloat[]{
  \includegraphics[width=.8\textwidth]{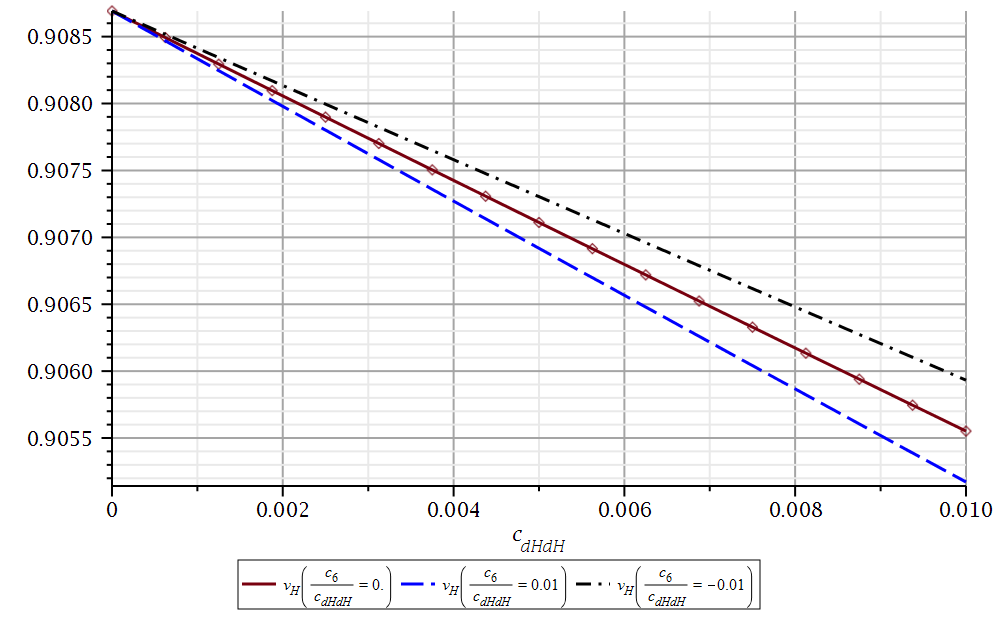}
\label{fig10b}
}
\caption{(a) The action calculated on EOM and (b) the value of the $h(\tau)$ field at $\tau =0$ for various $c_{dHdH}$ and $c_{6}$. 
The remaining parameters are $\bar{M}_{Pl}^2 = 10^2, 
\lambda_{H} = 6.0, m_{H}^2 = 0.2, a_{3} = -0.4, \xi_{H} = 0, c_{0} = 0.0$.
}
\label{fig10}
\end{figure}

After discussing the influence of the gravity mediated operators on the vacuum stability we will now focus on the non-geometric ones.
To this end, we discuss the case of dimension six operators $c_{6} h^6$ and much less often discussed 
$c_{dHdH} \nabla_{\mu}h^2 \nabla^{\mu} h^2$ one. 
After inspecting (\ref{cdHdH}) and (\ref{c6}) we may see that in our toy model these coefficients are related to each other 
at high energy. On the other hand, if we take into account running of these couplings their values in the low energy limit 
will differ, yet we will not discuss this running in this article. Having this in mind we plotted the influence of both operators 
on the vacuum stability in Figure~\ref{fig10a}. From it we may see that an increase in these coefficients leads to an increase 
of the stability of false vacuum. Moreover, if both coefficients in front of these operators are positive, this stability is enhanced 
in comparison to the case $c_6 =0$. On the other hand, in the case when $c_{6}$ is negative this vacuum stabilization is suppressed.  
In Figure~\ref{fig10b} we plotted the dependence of $v_H$ on the same coefficients. The value of the 
scalar field $h$ at $\tau=0$ decreases as we increase the $c_{dHdH}$ coefficient, this is also the case if we consider
variation in the $c_{6}$ coefficient. This decrease is weaker in the case when $c_{6}$ has an opposite sign to the $c_{dHdH}$ coefficient.

\begin{figure}[h]
\centering
\subfloat[]{
  \includegraphics[width=.8\textwidth]{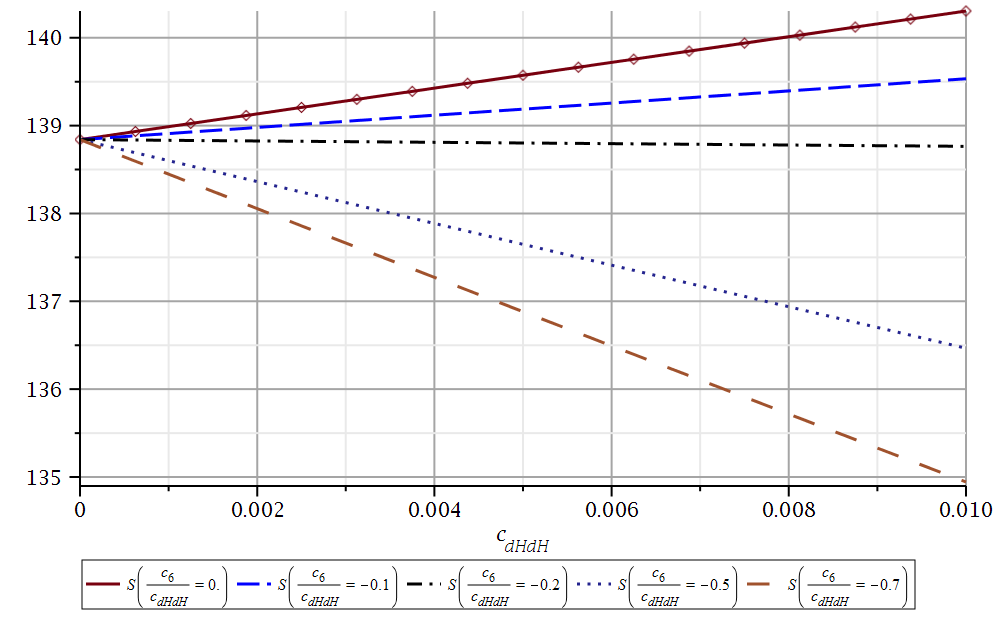}
\label{fig11a}
}
\quad
\subfloat[]{
  \includegraphics[width=.8\textwidth]{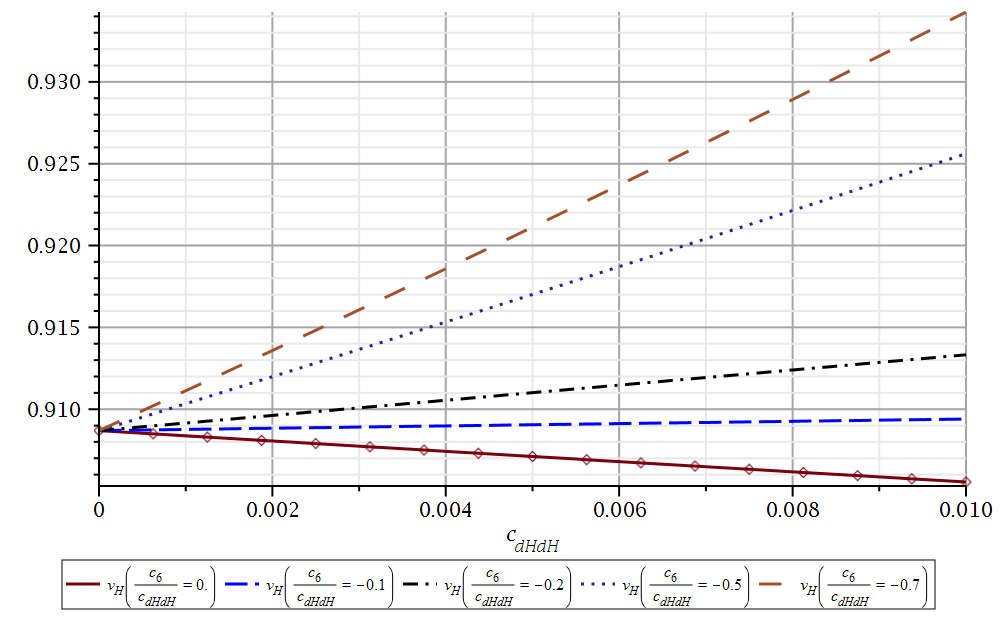}
\label{fig11b}
}
\caption{(a) The action calculated on EOM and (b) the value of the $h(\tau)$ field at $\tau =0$ for various $c_{dHdH}$ and $c_{6}$. 
The remaining parameters are $\bar{M}_{Pl}^2 = 10^2, 
\lambda_{H} = 6.0, m_{H}^2 = 0.2, a_{3} = -0.4, \xi_{H} = 0, c_{0} = 0.0$.}
\label{fig11}
\end{figure}

Focusing a little bit more on these coefficients leads us to some additional interesting conclusions. In Figure~\ref{fig11a}
we plotted the action calculated for various $\frac{c_6}{c_{dHdH}}$ rates. Before we discuss this let us make one technical remark.
Most of these rates lay beyond the allowed parameter space for our model. Nevertheless, they are physically reasonable in the 
sense that in the case when $c_6$ and $c_{dHdH}$ are treated as independent coefficients the rates represent physically allowed values for them.
The above situation could be the case when the integrated heavy sector was more complicated than the one containing a single massive scalar field.
Returning to Figure~\ref{fig11a}, if the $\frac{c_6}{c_{dHdH}}$ ratio is negative and big enough, 
the $c_6 h^6$ operators start to dominate and lead to a decrease in the action value and therefore to a decrease in 
the false vacuum lifetime. In our case the negativity of the aforementioned ratio implies that the coefficient $c_6$ is negative. This implies that
in a Lorentzian spacetime it will have an opposite sign to the quartic self-interaction term in the scalar field potential. This increase
in vacuum instability is in agreement with what we would expect by analyzing only the potential for the scalar field. 
By doing this we could write $\lambda h^4 - |c_6| h^6 = \lambda(h)_{eff}h^4$ and conclude that the $c_6$ term will make 
$\lambda(h)_{eff}$ negative for large enough fields therefore it would worsen the stability. 
For completeness, in Figure~\ref{fig11b} we plotted the behavior of $v_{H}$ for various $\frac{c_6}{c_{dHdH}}$
rates. For small rates $v_H$ decreases for an increasing $c_{dHdH}$ and for sufficiently large and negative $c_6$
this tendency is reversed.  
At this point let us note that the obtained type of behavior of the vacuum decay on $c_6$, namely an increase in stability for
positive $c_6$ and a decrease for the negative one is in agreement with previous results 
\cite{PhysRevLett.111.241801,Lalak:2014qua,PhysRevD.91.013003,doi:10.1142/S0217732315501898,Burda:2016mou}.

\subsection{Essential features in the language of the $UV$ theory coefficients}

In this subsection we reexpress our findings in terms of the $UV$ theory parameters $\xi_X$ and $\lambda_{HX}$.
Since we discussed various aspects of our results quite extensively in the previous subsection, here we will focus 
on the main problem, namely the influence of the aforementioned coefficients on the probability
of the false vacuum decay.   
Firstly, let us write again the cEFT parameters relevant for the research presented in this article, they are given by 
\begin{align}
\label{coefUV1}
c_{dHdH} &= \frac{\lambda_{HX}^2}{12 (4 \pi)^2}, \\
\label{coefUV2}
c_{HH} &= \frac{3 \lambda_{HX}^2}{12 (4 \pi)^2}(2\xi_x - \frac{1}{6}), \\
\label{coefUV3}
c_6 &= \frac{ \lambda_{HX}^3}{12 (4 \pi)^2}.
\end{align}  
As we may see from the above relations, the three cEFT coefficients depend on the two $UV$ ones.
In agreement with what was written in previous subsection we treated $c_6$ and $c_{dHdH}$ as connected 
to each other and $c_{HH}$ as independent of them.

\begin{figure}[h]
\includegraphics[width=.9\textwidth]{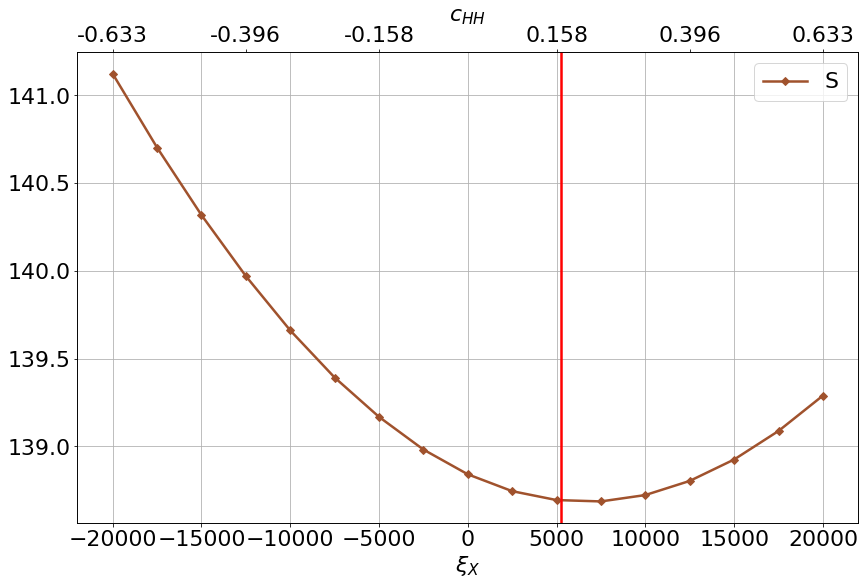}
\caption{The influence of $\xi_X$ on the false vacuum decay exponent. The remaining parameters are $\bar{M}_{Pl}^2 = 10^2, \lambda_{H} = 6.0, m_{H}^2 = 0.2, a_{3} = -0.4,  c_{0} = 0.0$ and 
$\lambda_{HX} = 0.1, c_{HH} = \frac{3 \lambda_{HX}^2}{12 (4 \pi)^2}(2\xi_x - \frac{1}{6}), 
c_{6} = \frac{\lambda_{HX}^3}{12 (4\pi)^2}, c_{dHdH} = \frac{\lambda_{HX}^2}{12 (4 \pi)^2}$. }
\label{fig12}
\end{figure}

In Figure \ref{fig12} we plotted the effect of the varying heavy scalar non-minimal coupling parameter $\xi_X$
on the on-shell value of the effective action.  For comparison, on the upper axis we also ploted the values of $c_{HH}$
as calculated for $\lambda_{HX} = 0.1$.  For a better comparison with Figure \ref{fig5a} we also plotted
a red vertical line representing $c_{HH} = \frac{1}{6}$, which corresponds to $\xi_X = 5263.87$. Firstly, we may see that the minimum of the on-shell effective action is located for large positive value of $\xi_X \approx 7500$. Secondly, a departure 
from that value of $\xi_X$ leads to an increase of $\mathcal{S}$ and therefore the stabilization of the false vacuum.

\begin{figure}[h]
\includegraphics[width=.9\textwidth]{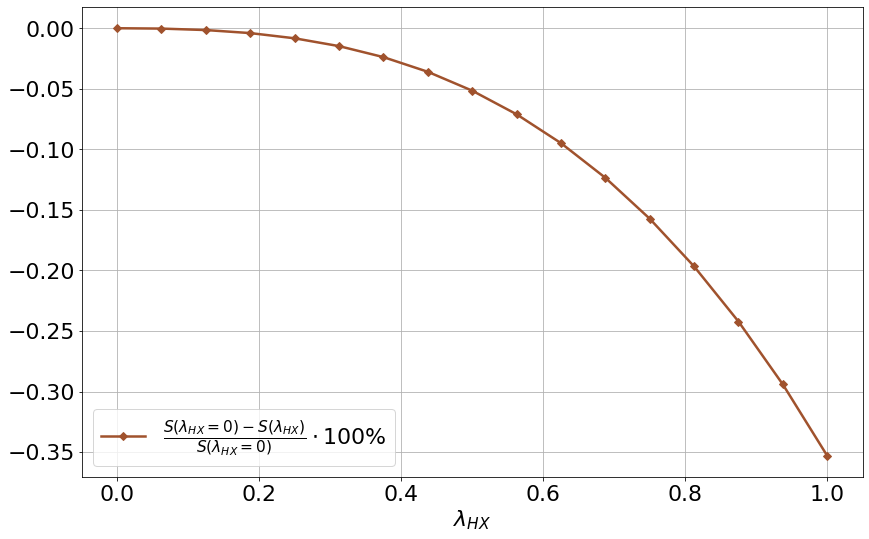}
\caption{The influence of $\lambda_{HX}$ on the false vacuum decay exponent. The remaining parameters are $\bar{M}_{Pl}^2 = 10^2, \lambda_{H} = 6.0, m_{H}^2 = 0.2, a_{3} = -0.4,  c_{0} = 0.0$ and 
$\xi_{X} = -1, c_{HH} = \frac{3 \lambda_{HX}^2}{12 (4 \pi)^2}(2\xi_x - \frac{1}{6}), 
c_{6} = \frac{\lambda_{HX}^3}{12 (4\pi)^2}, c_{dHdH} = \frac{\lambda_{HX}^2}{12 (4 \pi)^2}$. }
\label{fig13}
\end{figure}

In Figure \ref{fig13} we plotted the influence of the quartic portal coupling between the light and heavy sectors,
for this plot we assumed $\xi_X = -1$. It is worth noting that this coupling enters the coefficients of all considered higher dimensional operators, therefore $\lambda_{HX} = 0$ represents the action without any higher dimensional operators. 
To better display the effects of $\lambda_{HX}$ on the false vacuum decay we plotted the difference between the action calculated at $\lambda_{HX} =0$ and $\lambda_{HX} \neq 0$ normalized at $\lambda_{HX} = 0$ and expressed in percents. 
We see that as the portal coupling grows the difference becomes more negative which implies the stabilization of the false vacuum.
On the other hand, this difference even for $\lambda_{HX} \sim \mathcal{O}(1)$ is on a subpercent level.

\section{Summary}
\label{sec:summary}

In the article we discussed a possibility of using the curved spacetime Effective Field Theory (cEFT) to tackle the problem of 
investigating the influence of gravity on vacuum stability. To this end, we firstly obtained cEFT for the light scalar 
by integrating out the heavy one. Then we focused on investigating the role of the dimension six operators 
that contribute both to the potential and kinetic parts of the scalar field action. We also discussed the influence of the
novel higher dimensional gravity mediated operator that does not have its counterpart in the flat spacetime case. 
To aid our analysis we also used two observables, one of which was the Ricci scalar and was related to geometry 
and the second one was the Euclidean counterpart of energy density.   

For the non-minimal coupling of the light scalar to gravity we found that the value of the Euclidean action calculated at the 
solutions to the equations of motion changes monotonically with the change of $\xi_H$ for the $\xi_H \leqslant 0$ case.
Moreover, for this range of the non-minimal coupling it becomes higher as energy density inside the true vacuum bubble increases. 
This in turn implies that it is more costly to create such a bubble which is reflected in a bigger decay exponent, and therefore smaller
probability of false vacuum decay. 
An additional feature of the $\mathcal{S}(\xi_H)$ dependence we found is the presence of the minimum of $\mathcal{S}$ 
for $\xi_H$ slightly below $\xi_H = \frac{1}{6}$.

As far as $c_{HH}$ is concerned, we see a similar type of behavior. Namely, we see an improvement of vacuum stability for
$c_{HH} \leqslant 0$ and a minimum of $\mathcal{S}(c_{HH})$ for $c_{HH}$ slightly above $c_{HH} = \frac{1}{6}$.
This position of the minimum above $c_{HH} = \frac{1}{6}$ is a subtle difference as compared to the $\mathcal{S}(\xi_H)$
case. Moreover, we found out that for configurations with $(\xi_H \neq 0, c_{HH}=0)$ the pressure inside a true vacuum bubble
decays slower than for $(\xi_H = 0, c_{HH} \neq 0)$. This with the observation that the region of the energy density 
bigger than zero is wider for configurations of the first kind may lead us to the conclusion that $c_{HH} \neq 0$
leads to the bubbles of smaller width as compared to the configurations with equal in value $\xi_H$.

Let us also note that our investigation allows to observe some interesting features of the Ricci scalar rarely discussed 
in the context of vacuum stability. These were the non-monotonicity of $R$ as a function of Euclidean time and
the fact that the Ricci scalar changes its sign close to the bubble wall location. 
We found that the  region for which $R$ attains a positive value is also the region 
where the pressure associated with scalar field becomes negative.  

As for the operators of dimension six that are also present in the flat spacetime case we discussed the standard contribution to the scalar 
field potential with coefficient $c_6$ and much less often discussed contribution to the kinetic term with coefficient $c_{dHdH}$. 
In our cEFT approach they turn out to be linked to each other since both of them stem 
from the Higgs portal like coupling among scalar fields $\lambda_{HX}$ by the formula $\frac{c_6}{c_{dHdH}} = \lambda_{HX}$. 
From the analyzed cases we observed that if the abovementioned coefficients ratio is close to unity, the $c_6 h^6$ term will 
give a dominating contribution to the action. 
Moreover, if both $c_6$ and $c_{dHdH}$ are positive we observe an increase in the false vacuum stability which is in agreement 
with the flat space results for $c_6$.

In conclusion, we may state that applying cEFT to the case at hand turns out to be very productive and
revealed some novel features in the problem of vacuum stability in curved spacetime.


\section*{Acknowledgements}

The work of Z.L. has been partially supported by National Science Centre, Poland OPUS project 2017/27/B/ST2/02531.
A.N. was supported by the National Science Centre, Poland under a postdoctoral scholarship DEC-2016/20/S/ST2/00368.
{\L}.N. was supported by the National Science Centre, Poland under a grant DEC-2017/26/D/ST2/00193.




\bibliographystyle{JHEP}
\bibliography{cEFT_2s.bib}



\end{document}